\documentclass[aps,prb,superscriptaddress]{revtex4}
\usepackage{subfigure}
\usepackage{amsmath}
\usepackage{amssymb}
\usepackage{graphics}
\usepackage{rotating}
\usepackage{array}
\usepackage{graphicx}
\usepackage{hyperref}
\usepackage{xcolor}

\begin{document}

\title{Calculation of reaction constants using Transition Path Sampling with a local Lyapunov bias}
\author {Massimiliano Picciani}
\affiliation{CEA, DEN, Service de Recherches de M\'etallurgie Physique, F-91191 Gif-sur-Yvette, France}

\renewcommand{\baselinestretch}{1.2}\normalsize

\begin{abstract}

We propose an efficient method to compute reaction rate constants of thermally activated processes occurring in many-body systems at finite temperature. The method consists in two steps: first, trajectories are sampled using a transition path sampling (TPS) algorithm supplemented with a local Lyapunov bias favoring diverging trajectories. This enhances the probability of observing rare reactive trajectories between stable states during relatively short simulations.  Secondly, reaction constants are eventually estimated from the unbiased fraction of reactive trajectories, yielded by an appropriate statistical data analysis tool, the multistate Bennett acceptance ratio (MBAR) package. 

In order to test our algorithm, we compute reaction constants for structural transitions in LJ$_{38}$, a well studied Lennard-Jones cluster, comparing our results to values previously reported in the literature. Additionally, we apply our method to the calculation of reaction rates for the migration of vacancies in an $\alpha$-Iron crystal, for temperatures ranging from 300 K to 850 K. Vacancy diffusion rates associated with activation barriers at finite temperature are then evaluated, and shown to be substantially different from values obtained using the standard harmonic approximation.

\end{abstract}

\maketitle

\section{Introduction}

Rare events are physical events occurring with low probability in numerous many-body systems, encompassing a wide range of phenomena, from condensed matter physics to proteins: they often involve thermally activated processes, i.e. passages between different stable configurations of the system that are separated by free energy barriers. Evaluating  the frequency of these rare events from numerical simulations can be very time consuming, as the probability of observing one of these passages is very low: for instance, the migration of a vacancy in $\alpha$-Iron at 500K typically happens every microsecond, while the usual time steps for molecular dynamics is of the order of a few femtoseconds. 
In the last decades, several techniques have been developed in order to accelerate the dynamics, and enhance the probability of observing infrequent events during a short simulation, based on importance sampling.~\cite{W2003,BCDG2002,MB1998,picciani2011simulating} 

Among these techniques, transition path sampling~\cite{BCDG2002} (TPS) allows to estimate the frequency of rare events by means of path ensemble averages of physical observables: reaction rates for appropriate time scales are indeed evaluated from the ratio of the number of reactive trajectories on the total amount of paths sampled. However, a sufficient number of reactive paths has to be observed, in order to obtain a reliable statistics.  

Herein, we propose a transition path sampling algorithm where the fraction of reactive paths sampled is enhanced using an adequate bias that favors diverging trajectories. It is indeed possible to show~\cite{TK2006,geiger2010identifying} that reactive paths we want to sample share important features with diverging trajectories observed in chaotic systems. Therefore, a suitable parameter that quantifies the divergence of dynamic trajectories can be exploited also to bias a path sampling algorithm aimed to reproduce reactive paths.

The main instrument proposed in the literature to quantify chaotic behavior of dynamical systems is the evaluation of Lyapunov exponents,~\cite{ott2002chaos} that are usually employed to estimate the sensitivity of deterministic systems to small changes in initial conditions. For this feature, they have been widely studied,~\cite{lichtenberg1992regular} both analytically and numerically, in hamiltonian as well as in nonlinear systems of small dimensions. Moreover, the use of Lyapunov exponents to characterize numerically phase transitions in finite size systems has been extensively explored in the past years, and many noticeable results have been obtained in the early 90's.~\cite{hinde1992chaos,hinde1993chaotic,calvo1998chaos,amitrano1992probability} 

Resorting to Lyapunov exponents in order to achieve numerical ergodicity and localize saddles and transition paths has been done recently with the Lyapunov-weighted dynamics method proposed by Tailleur and Kurchan:~\cite{TK2006} in this sampling scheme, a set of clones are copied or deleted depending on a probability weight computed from quantities related to Lyapunov exponents. After this work, the paper of Geiger and Dellago~\cite{geiger2010identifying} have shown how to couple the chaoticity features of a dynamical system to a TPS technique for sampling deterministic trajectories, using an indicator for diverging trajectories borrowed from studies on planetary systems~\cite{sandor2004relative}, the relative Lyapunov indicator (RLI).

In this paper, we present a chaoticity indicator different from RLI, and based on local Lyapunov numbers, that are quantities closely related to Lyapunov exponents. This indicator is used to introduce a bias in the path sampling scheme, thus obtaining a Lyapunov biased TPS (LyTPS) method that will be in the sequel applied to complex many-body systems, like the well known optimization benchmark model LJ$_{38}$. Furthermore, we show how reaction rate constants can be recovered from biased TPS quantities, resorting to an appropriate statistical analysis to unbias reaction rate values computed in a LyTPS framework.

This paper is organized as follows. In Sec.~\ref{sec:Lyapunov-Exponents-in}, we first recall the basic concepts of Lyapunov exponents for dynamical systems. We will then briefly review the use that has been made of them in numerical algorithms to characterize phase transitions, or in importance sampling contexts. We then expose how to use local Lyapunov numbers in the context of a Transition Path Sampling to determine saddle points and reactive paths (Sec.~\ref{sec:Transition-path-sampling}). Reaction constants are computed from the fraction of reactive paths using the Bennett-Chandler approach \cite{chandler1987introduction} of population correlation functions and the standard TPS technique \cite{DBCD1998}; we also explain how unbiased reaction constants are recovered from a Lyapunov biased algorithm thanks to the multistate Bennett acceptance ratio~\cite{shirts2008statistically} (MBAR) method (Sec.~\ref{sec:Reaction-rate-constants-calculation}). Finally, numerical results concerning  the application of our method to solid-solid structural transition in LJ$_{38}$ and vacancy migration in $\alpha$-Iron are presented (Sec.~\ref{Numerical results}).

\section{{\normalsize Lyapunov Exponents in dynamical systems \label{sec:Lyapunov-Exponents-in}}}

We briefly present here the theory of Lyapunov exponents, mainly following Ott \cite{ott2002chaos}; then we propose a formulation allowing the use of these exponents in importance sampling techniques. 

\subsection{Discrete dynamics and numerical applications: state of the art}\label{rliecc}

In numerical applications, dynamics are discrete: the evolution of state vector $\mathbf{x}$ at time step $n$ is described by a \emph{mapping} $\mathbf{x}_{n+1}=\mathbf{M}(\mathbf{x}_{n})$, where $\mathbf{M}$ is a matrix expressing the system evolution from one time step to the following. 

Let the system be at time $t=0$ in an initial position $\mathbf{x}_{0}$,
and let $\delta\mathbf{x}_{0}$ be a small perturbation applied to
this initial state. The dynamics of such a perturbed system can then
be denoted using a new state vector $\tilde{\mathbf{x}}=\mathbf{x}+\delta\mathbf{x}$,
whose time evolution will be \begin{equation}
{\tilde{\mathbf{x}}_{n+1}}={\mathbf{x}}_{n+1}+{\delta\mathbf{x}}_{n+1}=\mathbf{M}(\mathbf{\tilde{x}}_{n})=\mathbf{M}(\mathbf{x}_{n}+\delta\mathbf{x}_{n})\label{eq:1}\end{equation}
For a sufficiently small perturbation $\delta\mathbf{x}$, it is possible to linearize
the map $\mathbf{M}$ as \begin{equation}
\mathbf{M}(\mathbf{x}_{n}+\delta\mathbf{x}_{n})=\mathbf{M}(\mathbf{x}_{n})+D\mathbf{M}(\mathbf{x}_{n})\cdot\delta\mathbf{x}_{n}+O(\delta\mathbf{x}_{n}^{2})\label{eq:2b}\end{equation}
The time evolution of a small perturbation to the initial state vector then reads~\cite{ott2002chaos}
\begin{equation}
\delta\mathbf{x}_{n+1}=D\mathbf{M}\mathbf{(}\mathbf{x}_{n})\cdot\delta\mathbf{x}_{n}\label{eq:7}
\end{equation}
where $D\mathbf{M}(\mathbf{x}_{n})$ is the Jacobian matrix of the map $\mathbf{M}$. Inserting particular solutions $\delta\mathbf{x}_{n}=\mathbf{e}[\Lambda]^{n}$ in Eq.~\eqref{eq:7}, one finds an eigenvalue equation 
\begin{equation}\label{eq.eig}
D\mathbf{M}\mathbf{(}\mathbf{x}_{n})\cdot\mathbf{e}=\Lambda\cdot\mathbf{e}.
\end{equation}
In the discrete case, the eigenvalues $\Lambda_k$ of $D\mathbf{M}\mathbf{(}\mathbf{x}_{n})$, solutions of Eq.~\eqref{eq.eig}, are called \emph{Lyapunov numbers} rather the Lyapunov exponents, and trajectories are \emph{unstable} for $\left|\Lambda_{k}\right|>1,$ and stable otherwise. The largest eigenvalue $\Lambda^{MAX}$ will be associated to the eigenvector $\mathbf{e}^{MAX}$ indicating the direction of maximal growth of the perturbation $\delta\mathbf{x}$. We introduce the matricial product 
\begin{equation}
D\mathbf{M}^{n}\mathbf{(}\mathbf{x}_{0})=D\mathbf{M}\mathbf{(}\mathbf{x}_{n-1})\cdots D\mathbf{M}(\mathbf{x}_{0})\label{eq:12}\end{equation} between the Jacobian matrices of the hamiltonian map at successive time steps, and we express the perturbation at time step \emph{n} with respect to the initial perturbation $\delta\mathbf{x}_{0}$ as \begin{equation}
\delta\mathbf{x}_{n}=D\mathbf{M}^{n}\mathbf{(}\mathbf{x}_{0})\cdot\delta\mathbf{x}_{0}\end{equation}
Defining with $\left\Vert\delta\mathbf{x}_{0}\right\Vert$ the Euclidean norm of $\delta\mathbf{x}_{0}$ in phase space, the \emph{Lyapunov exponents} $h$, given the initial condition $\mathbf{x}_{0}$ and the initial perturbation orientation $\mathbf{u}_{0}=\delta\mathbf{x}_{0}/\left\Vert\delta\mathbf{x}_{0}\right\Vert$ are 
\begin{equation} \label{eq:lyap_exp}
h(\mathbf{x}_{0},\mathbf{u}_{0})=\lim_{n\rightarrow\infty}\frac{1}{n}\ln\left(\frac{\left\Vert\delta\mathbf{x}_{n}\right\Vert}{\left\Vert\delta\mathbf{x}_{0}\right\Vert}\right)=\lim_{n\rightarrow\infty}\frac{1}{n}\ln\left\Vert D\mathbf{M}^{n}\mathbf{(}\mathbf{x}_{0})\cdot\mathbf{u}_{0}\right\Vert
\end{equation}
For \emph{N}-body  hamiltonian systems, having \emph{3N} degrees of freedom, $\mathbf{x}(t)$ is a \emph{6N}-dimensional state vector accounting for both the positions and momenta of the $N$ particles. For the \emph{6N}-dimensional hamiltonian map there will be \emph{6N} Lyapunov exponents, usually ordered in literature from the largest to the smallest ($h_{1}\geq\cdots\geq h_{6N}$). As stated in Ref.~\cite{ott2002chaos}, Oseledec's multiplicative ergodic theorem \cite{oseledec1968multiplicative} guarantees the existence of the limits used in the definition of the Lyapunov exponents in Eq.~\eqref{eq:lyap_exp} under very general circumstances.

The Lyapunov exponents are related to the aforementioned Lyapunov numbers as $\Lambda_{k}=\exp\left[h_{k}\right]$.  From Eq.~\eqref{eq:lyap_exp}, we can also define \emph{finite-time Lyapunov exponents}:
\begin{equation}\label{finite-time_lyap}
\bar{h}_{n}(\mathbf{x}_{0},\mathbf{u}_{0})=\frac{1}{n}\ln\left(\frac{\left\Vert \delta\mathbf{x}_{n}\right\Vert}{\left\Vert \delta\mathbf{x}_{0}\right\Vert}\right)=\frac{1}{n}\ln\left\Vert D\mathbf{M}^{n}\mathbf{(}\mathbf{x}_{0})\cdot\mathbf{u}_{0}\right\Vert
\end{equation}
For long enough times, the greatest Lyapunov number $\Lambda_1$ will give the dominant contribution to the perturbation evolution, and the associated eigenvector $\mathbf{e}_1$ indicates the direction of maximum growth of the perturbation $\delta\mathbf{x}$.

We stress here that finite-time Lyapunov exponents are calculated for a given $\mathbf{x}_{0}$ and that, strictly speaking, their values do depend on the initial orientation $\mathbf{u}_{0}$. It is however shown that the largest exponent $h_{1}(\mathbf{x}_{0},\mathbf{u}_{0})$ is approximately independent of the choice of $\mathbf{u}_{0}$ in Hamiltonian ergodic systems\cite{ott2002chaos,meyer}, while the complete spectra of  finite-time exponents can be determined using specific numerical techniques.~\cite{benettin1980lyapunov} 

In numerical simulations, only finite-time Lyapunov exponents can be estimated, due to limited amount of CPU time. To evaluate $\bar{h}_{n}$ from Eq.~\eqref{finite-time_lyap} we should compute the matricial product of Eq.~\eqref{eq:12}. For systems with many degrees of freedom, a calculation of this matrix product is not possible analytically, and is numerically expensive. 
The solution most followed in literature consists in directly evaluating the quantity 
\begin{equation}\label{evalh0}
\bar{h}_{n}(\mathbf{x}_{0},\mathbf{u}_{0})=\frac{1}{n}\ln\left(\frac{\left\Vert\delta\mathbf{x}_{n}\right\Vert}{\left\Vert\delta\mathbf{x}_{0}\right\Vert}\right)
\end{equation}
given by the distance $\left\Vert\delta\mathbf{x}_{n}\right\Vert$ between two nearby dynamical trajectories, the first one started from the initial state $\mathbf{x}_{0}$ and the second one from the perturbed configuration $\tilde{\mathbf{x}}_{0}=\mathbf{x}_{0}+\delta\mathbf{x}_{0}$ after \emph{n} time steps. 

The use of Eq.~\eqref{evalh0} as a mean to evaluate finite-time Lyapunov exponents has two main drawbacks: the need of computing two trajectories to evaluate a single Lyapunov exponent, thus doubling computational cost, and the fact that values obtained for $\bar{h}$ can be sensitive to initial conditions, because of   the dependence of the computed finite-time Lyapunov exponents from the choice of the orientation of initial perturbation $\mathbf{u}_{0}$, as recalled above.

Several numerical strategies have been proposed to bypass these two problems. The tangent space method \cite{benettin1980lyapunov, shimada,dellagohooverposch} assigns to each state $\mathbf x_t$ of the trajectory started in $\mathbf x_0$ a vector $\mathbf{u}(\mathbf{x}_t)$. These vectors are computed from the local hessian matrix of the hamiltonian mapping, and their norms indicate the distance between the current trajectory and the perturbed one, i.e. $\mathbf{u}(t) \sim \delta \mathbf{x}(t)$. As these distances evolve exponentially (see Eq.~\eqref{eq:lyap_exp}), the lengths of the vectors $\mathbf u$ can quickly diverge or vanish: a reorthonormalization of $\mathbf u$ at each time step is therefore required, for instance with a Gram-Smith algorithm. This method has been implemented in the literature~\cite{lichtenberg1992regular}, for instance in the context of Lyapunov weighted dynamics.~\cite{TK2004,TTK2006,TK2006} To make this algorithm independent of the choice of the first vector $\mathbf{u}(\mathbf{x}_0)$, one could integrate the equations of motion backward in time from $\mathbf x_0$ for a duration $\tau$, and then reintegrate the evolution of $\mathbf{u}(\mathbf{x}_t)$ forward until $t=0$(see Ref.~\cite{dellagohooverposch}). In this way, $\mathbf{u}(\mathbf{x}_0)$ would be automatically oriented in the direction of maximum growth. However, the duration $\tau$ should be long enough to ensure the loss of correlation between the orientation of $\mathbf{u}(\mathbf{x}_{-\tau})$ and $\mathbf{u}(\mathbf{x}_0)$, thus requiring the computation of long trajectories at sustained computational cost.~\cite{geiger2010identifying,dellagohooverposch} 

Another evaluation of finite-time Lyapunov exponents can be obtained with the Relative Lyapunov Indicator (RLI), elaborated by S\`andor
\emph{et al.}~\cite{sandor2004relative} in the context of planetary
trajectories, and further used in a Lyapunov weighted path
sampling scheme~\cite{geiger2010identifying}. The main idea is to
compare finite-time Lyapunov exponents $\bar{h}$ for trajectories
starting very close, say in $\mathbf{x}_{0}$ and $\mathbf{x}_{0}+\Delta\mathbf{x}_{0}$. 
The difference between finite-time exponents at time step $n$ can be written as 
\begin{equation}
\Delta\bar{h}_{n}(\mathbf{x}_{0},\mathbf{u}_{0})=\left|\bar{h}_{n}(\mathbf{x}_{0}+\Delta\mathbf{x}_{0},\mathbf{u}_{0})-\bar{h}_{n}(\mathbf{x}_{0},\mathbf{u}_{0})\right|                                                \end{equation}
and will in general undergo strong fluctuation,~\cite{sandor2004relative} which can be smoothed
by an average over $\mathcal N$ trajectory steps: in this way one defines
the RLI as the quantity 
\begin{equation}
R(\mathbf{x}_{0},\mathbf{u}_{0})=\frac{1}{\mathcal N}\sum_{i=1}^{\mathcal N}\Delta\bar{h}_{i}(\mathbf{x}_{0},\mathbf{u}_{0})
\end{equation}
This average over the entire trajectory length reduces \cite{sandor2004relative} the dependence of the computed finite-time Lyapunov exponents on the orientation of initial perturbation, but introduces an additional dependence on $\Delta\mathbf{x}_{0}$. Both finite-time Lyapunov exponents are calculated evaluating the distance between two trajectories evolving close to each other (instead of the matricial product of Eq.~\eqref{eq:12}): in terms of computational cost, four trajectories are computed to obtain a single RLI. 

In the following, we propose a faster and orientation-independent way to evaluate the divergence of hamiltonian trajectories, alternative to tangent space method and RLI, to be used in the path-sampling scheme described in Sec.~\ref{sec:Transition-path-sampling}.


\subsection{{\normalsize Hamiltonian dynamics \label{sub:Dynamical-systems-with}}}

We restrict our focus to systems with deterministic dynamics governed by an hamiltonian $\mathcal H=\sum_{i=1}^{N}\frac{\mathbf{p}_{i}^{2}}{2m}+V(\mathbf{q})$. The time evolution of the state vector $\mathbf{x}=(\mathbf{q},\mathbf{p})$ (also indicated as ``hamiltonian flow'' \cite{lichtenberg1992regular}) directly follows
from Hamilton equations, \begin{equation}
\mathbf{\dot{x}}=\left(\begin{array}{cc}
\mathbf{0} & \mathbf{I}_{3N}\\
-\mathbf{I}_{3N} & \mathbf{0}\end{array}\right)\left(\begin{array}{c}
\frac{\partial \mathcal H(\mathbf{q},\mathbf{p)}}{\partial\mathbf{q}}\\
\frac{\partial \mathcal H(\mathbf{q},\mathbf{p})}{\partial\mathbf{p}}\end{array}\right)\label{eq:6}\end{equation}
The evolution of a small perturbation of positions and momenta for a standard hamiltonian is then given by 
\begin{equation}
\delta\mathbf{\dot{x}}=\left(\begin{array}{cc}
\mathbf{0} & \mathbf{G}\\
-\mathbf{\frac{\partial^{2}V(\mathbf{q})}{\partial\mathbf{q}\partial\mathbf{q}}} & \mathbf{0}\end{array}\right)\delta\mathbf{x}\end{equation}
where $\mathbf G$ is the $3N\times3N$ inverse mass matrix of elements ${G}_{ij}=\delta_{ij}/m_i$.

Let us discretize the hamiltonian dynamics in Eq.~\eqref{eq:6} with the velocity Verlet algorithm:
\begin{eqnarray}\label{discr2}\nonumber
p_{i,n+1/2} & = & p_{i,n}-\frac{1}{2}dt\cdot\frac{\partial V(\mathbf{q}_n)}{\partial q_{i,n}}\\
q_{i,n+1} & = & q_{i,n}+\frac{dt}{m}p_{i,n+1/2}\\\nonumber
p_{i,n+1} & = & p_{i,n+1/2}-\frac{1}{2}dt\cdot\frac{\partial V(\mathbf{q}_{n+1})}{\partial q_{i,n+1}}.
\end{eqnarray} This algorithm is accurate to second order and numerically stable~\cite{frenkel2002understanding}. It will be used in Sec.~\ref{sec:Transition-path-sampling} to generate dynamical trajectories in numerical applications. The perturbation $\delta\mathbf{{x}}_{n+1}$ can be evaluated with respect to $\delta\mathbf{{x}}_{n}$ using Eq.~\eqref{eq:7}, where the jacobian matrix of the hamiltonian mapping for a velocity Verlet discretization reads~\cite{hinde1992chaos}
\begin{equation}\label{eq:secordjac}
D\mathbf{M}(\mathbf{x}_{n})=\left[\begin{array}{cc}
\mathbf{I}-\frac{dt^2}{2m}\mathbf{H}(\mathbf{x}_n) & \mathbf{G}\\
-\frac{dt}{2m}\left\{\mathbf{H}(\mathbf{x}_n)+\mathbf{H}(\mathbf{x}_{n+1})\left[\mathbf{I}-\frac{dt^2}{2}\mathbf{H}(\mathbf{x}_n)\right]\right\}& \mathbf{I}-\frac{dt^2}{2m}\mathbf{H}(\mathbf{x}_{n+1})\end{array}\right]
\end{equation}In the upper right and bottom left blocks we introduced the hessian matrix of the potential energy $\mathbf{H}$ at states $\mathbf{x}_{n}$ and $\mathbf{x}_{n+1}$, respectively.

The jacobian matrix of Eq.~\eqref{eq:secordjac} obtained with the velocity Verlet scheme of Eq.~\eqref{discr2} contains the hessian matrices at steps $n$ and $n+1$. Therefore, manipulating $D\mathbf{M}(\mathbf{x}_{n})$ is numerically expensive.  A simpler expression for the jacobian matrix can be obtained from the less accurate Euler discretization algorithm: Eq.~\eqref{eq:6} becomes a set of \emph{6N} coupled equations of motion 
\begin{eqnarray}\label{discr1}\nonumber
q_{i,n+1} & = & q_{i,n}+dt\cdot\frac{p_{i,n}}{m_{i}}\\
p_{i,n+1} & = & p_{i,n}-dt\cdot\frac{\partial V}{\partial q_{i,n}}\end{eqnarray}
where $i=1,...,3N$, so that the jacobian matrix of the hamiltonian map reads 
\begin{equation}\label{eq:dm_hamilt}
D\mathbf{M}(\mathbf{x}_{n})=\left[\mathbf{I}_{6N}+dt\left(\begin{array}{cc}
\mathbf{0} & \mathbf{G}\\
-\mathbf{H}(\mathbf{x}_n) & \mathbf{0}\end{array}\right)\right].\end{equation} Again, the perturbation $\delta \mathbf{x}_{n}$ at each time step can be evaluated by inserting Eq.~\eqref{eq:dm_hamilt} in Eq.~\eqref{eq:7}.

The difference between the jacobian matrix $D\mathbf{M}(\mathbf{x}_{n})$ of Eq.~\eqref{eq:dm_hamilt} and the one of Eq.~\eqref{eq:secordjac} consists in second order terms. However, it is numerically less expensive to manipulate the former than the latter, as $D\mathbf{M}(\mathbf{x}_{n})$ of Eq.~\eqref{eq:dm_hamilt} requires to evaluate the hessian only at time step $n$. In the following, we will be interested  in computing the eigenvalues of $D\mathbf{M}(\mathbf{x}_{n})$, in order to obtain a bias favoring reactive trajectories: this bias will be removed at the end, so it would be useless to spend CPU time to accurately evaluate the jacobian matrix. Therefore, accordingly to Ref.~\cite{hinde1992chaos}, we consider the Euler scheme (Eq.~\eqref{discr1}) precise enough for our purposes, and we use the first order Euler discretization of Eq.~\eqref{eq:dm_hamilt} to compute $D\mathbf{M}(\mathbf{x}_{n})$.

\subsection{{\normalsize Maximum local Lyapunov numbers \label{sub:Maximum-Lyapunov-exponents}}}

Using the discretized hamiltonian dynamics given in Eq.~\eqref{discr1}, we proceed by computing at each time step the maximum local Lyapunov number~\cite{abarbanel1992}, given by the largest eigenvalue of $D\mathbf{M}(\mathbf{x}_n)$ (Eq.~\eqref{eq:dm_hamilt}). 
The $6N$ eigenvalues $\Lambda_n$ of $D\mathbf{M}(\mathbf{x}_n)$, computed at time step $n$, can be obtained writing Eq.~\eqref{eq.eig} as a secular equation \begin{equation}\label{eq:det_lambda}
P(\Lambda_n)=det\left\{\Lambda_{n} \mathbf{I}_{6N}-D\mathbf{M}(\mathbf{x}_n)\right\}=0.
\end{equation}
whose solutions are $\emph{3N}$ pairs of eigenvalues $\Lambda_n$ of $D\mathbf{M}(\mathbf{x}_{n})$ , because of the simplectic properties of the hamiltonian mapping matrix $\mathbf{M}$.~\cite{ott2002chaos} These eigenvalues are given by the expression~\cite{hinde1992chaos}
\begin{equation} \label{eq:lambda_eigenv}
\Lambda_{j,n}^{\pm}=1\pm  dt\sqrt{-\lambda_{j,n}}\quad\forall j=1,\ldots,3N
\end{equation}
where the $\lambda_{j,n}$ correspond to the eigenvalues of the mass-weighted Hessian $\mathbf{H'}(\mathbf{x}_{n})$ of components
\begin{equation}
\label{eq:eigen}
\mathbf{H'}_{ij}(\mathbf{x}_{n})=\frac{1}{\sqrt{m_i m_j}} \frac{\partial^{2}V(\mathbf{x}_{n})}{\partial{q}_i\partial {q}_j} .
\end{equation} 

Eq.~\eqref{eq:lambda_eigenv} shows that at each time step $n$ the jacobian eigenvalues $\Lambda_{j,n}$, i.e. the local Lyapunov numbers, depend on the potential energy surface through the hessian eigenvalues $\lambda_{j,n}$: unstable configurations $\mathbf{x}_n$, such as saddle points are characterized by the occurrence of some negative $\lambda_{j,n}$, and correspond to real and positive local Lyapunov numbers $\Lambda_{j,n}$. Conversely, stable states have all $\lambda_{j,n}$  positive or imaginary local Lyapunov numbers with unitary real part. In the following path sampling scheme of Sec.~\ref{sec:Transition-path-sampling} we neglect the imaginary part of $\Lambda_{j,n}$ given by stable states. This is not an issue because it is sufficient to determine real and positive Lyapunov exponents to characterize unstable dynamics which we are interested in, and the imaginary part of the jacobian eigenvalues $\Lambda_{j,n}$ can be neglected. Indeed, the imaginary part of the Lyapunov exponents does not affect the stability of the system, but only indicates if $\delta\mathbf{x}_n$ is spiraling clockwise or counterclockwise.

At each time step $n$, the most negative eigenvalue of the hessian matrix $\lambda_n^{min}$ gives the eigenvalue of the Jacobian matrix $D\mathbf{M}(\mathbf{x}_{n})$ with the largest real part. We write the maximum local Lyapunov number as
\begin{equation}\label{maxlyap_ind}
\Lambda^{MAX}_{n}=1+dt\sqrt{\max(0,-\lambda_{n}^{min})}
\end{equation}
such that Eq.~\eqref{maxlyap_ind} entails $\Lambda^{MAX}_{n}=1$ for stable configurations, where all $\lambda$ are positive, and $\Lambda^{MAX}_{n}>1$ for unstable configurations having a negative spectra.  

Eigenvalues $\lambda_{n}$ are found by diagonalization of the hessian matrix $\mathbf H$. This 
can be computationally very expensive for systems with a large number of degrees of freedom. One efficient solution consists in extracting only the lowest eigenvalue $\lambda_{n}^{min}$ needed to evaluate $\Lambda_n^{MAX}$ using the Lanczos algorithm \cite{lanczos1961applied}. This iterative algorithm finds extremal eigenvalues of any matrix with a reduced computational cost, diagonalizing only a submatrix of the initial one (see for example appendix A of Ref.~\cite{marinica2011energy} for details). As pointed out in Ref.~\cite{cosminart}, a $15 \times 15$ Lanczos submatrix is sufficient to detect negative eigenvalues. Moreover, it is possible to decrease the submatrix size to as little as $4 \times 4$ by verifying at each iteration that the Lanczos solution is stable; if not, repeat the calculation until a the solution is converged. Hence, the most negative eigenvalue, corresponding to the most unstable direction of the potential energy surface at a give system position in the phase-space at a given instant can be extracted in a very few iterations. This is the numerical method we will apply in the following to evaluate $\Lambda_n^{MAX}$.

\subsection{{\normalsize Lyapunov indicator for dynamical trajectories \label{sub:Lyapunov-indic}}}

A dynamical trajectory is defined as an ordered sequence of states
in phase space separated by a small time increment $\delta t$, and denoted as 
$\mathbf{z}=\left\{ \mathbf{x}_{0},...,\mathbf{x}_{\tau}\right\} $, 
i.e. a path of total length $\tau$ composed by $\mathcal N=\frac{\tau}{\delta t}+1$
state vectors. We introduce a Lyapunov indicator for path $\mathbf z$
\begin{equation}
L(\mathbf z)=\frac{1}{\mathcal N}\ln\prod_{n=1}^{\mathcal N}\Lambda^{MAX}_n\label{eq:15}
\end{equation}
given by the average of the maximum Lyapunov number of Eq~\eqref{maxlyap_ind} over the whole trajectory.

The verification of the difference between finite-time Lyapunov exponents estimated by RLI or tangent space method and $L(\mathbf z)$  from Eq.~\eqref{eq:15} is beyond the scope of this article. We stress instead that the indicator proposed in Eq.~\eqref{eq:15}, being based on the hessian spectra $\lambda$, is strictly related to the topological properties of the potential energy surface, thus encodes local information on the stable or unstable configurations  sampled in phase space by a given trajectory. Henceforth, we consider this Lyapunov indicator suitable for importance sampling techniques.

The largest local Lyapunov number of Eq.~\eqref{maxlyap_ind} were evaluated by Hinde \textit{et al.}~\cite{hinde1992chaos} to study the dependence of the Kolmogorov entropy on the potential energy surface of small Lennard-Jones clusters. In that work, the Lyapunov exponents derived from the eigenvalues of the jacobian matrix of the hamiltonian mapping are summed over trajectories of different lengths, thus obtaining - using Pesin's theorem ~\cite{pesin1977characteristic} - an estimation of Kolmogorov entropy. Using Eq.~\eqref{maxlyap_ind} to evaluate finite-time Lyapunov exponents was proven to be a quite successfull approach, as it yields enough information to quantify the degree of instability of phase space trajectories. This supports the idea that evaluating the global Lyapunov exponent from local Lyapunov numbers allow to correctly characterize the chaotic properties of the system.

\section{Transition path sampling with a Lyapunov bias \label{sec:Transition-path-sampling}}

The idea of sampling the phase space of a many-body system through paths generated by molecular dynamics, using the Metropolis algorithm, has been  
developed by Dellago and coworkers.~\cite{DBCD1998,dellago1999calculation, dellago2002transition} This approach is called transition path sampling (TPS). 
In this present work, we introduce a bias in the TPS algorithm in order to favor the sampling of reactive trajectories. The bias is proportional to the Lyapunov indicator $L(\mathbf{z})$ obtained in Eq.~\eqref{eq:15}.

Each path $\mathbf{z}$ constrained to start in a reactant basin A is equipped with a probability density 
\begin{equation} 
\mathcal{P}^{\alpha}_{A}(\mathbf{z})=\frac{1}{Z^{\alpha}_{A}}\exp\left\{\alpha L(\mathbf{z})-\beta \mathrm H(\mathbf{z})\right\}\varphi_{A}^{\alpha}(\mathbf{x}_{0}) \label{eq:9}
\end{equation}
where $Z^{\alpha}_{A}$ is the partition function on the biased trajectory ensemble, and function $\varphi_{A}^{\alpha}(\mathbf{x}_{0})$ is an additional term linking the initial state $\mathbf{x}_{0}$ of the path to state $\mathbf{x}_{A}$. Different choices for $\varphi_{A}^{\alpha}$ are possible, for instance $\varphi_{A}^{\alpha}(\mathbf{x}_{0})=h_{A}(\mathbf{x}_{0})$, where $h_{A}$ is an indicator function on A, such that \begin{equation}\label{ha}
h_{A}(\mathbf{x})=\begin{cases}
1 & \mathbf{x} \in A\\
0 & \mathbf{x}\notin A \end{cases},\end{equation} or
\begin{equation}\label{costr_funct_1}
\varphi_{A}^{\alpha}(\mathbf{x}_{0})=\exp\left\{ -\frac{1}{2}\kappa^{\alpha}(\mathbf{x}_{0}-\mathbf{x}_{A})^{2}\right\} 
\end{equation}
that accounts for having a tunable spring of stiffness $\kappa^{\alpha}$ linking the origin of path $\mathbf{z}$ to state A. In this last case, the stiffness parameter $\kappa^{\alpha}$ can be tuned to counterbalance the strenght of the bias.
We denoted in Eq.~\eqref{eq:9} for simplicity
\begin{equation}\label{eqhz}
\exp\left\{-\beta\mathrm H(\mathbf{z})\right\}=\rho(\mathbf{x}_{0})\prod_{i=0}^{\tau/\delta t-1}\delta\left[\mathbf{x}_{(i+1)\delta t}-\phi_{\delta t}(\mathbf{x}_{i\delta t})\right] 
\end{equation}
as the dynamical path probability arising from the deterministic propagation of the trajectory. In Eq.~\eqref{eqhz} $\rho(\mathbf{x}_{0})=\exp(-\beta \mathcal H(\mathbf{x}_{0}))$ is the unnormalized canonical distribution at inverse temperature $\beta$ from which the initial configuration is selected, and $\phi$ is the \emph{temporal propagator} $\mathbf{x}_{t}=\phi_{t}(\mathbf{x}_{0})$ associated to the deterministic dynamics~\cite{dellago2002transition}.

In this biased ensemble, choosing positive $\alpha$ enhances the probability weights of trajectories with a large Lyapunov indicator $L(\mathbf{z})$, favors via Eq.~\eqref{eq:lambda_eigenv} the sampling of reactive paths passing over saddles and unstable directions of the potential energy landscape. On the contrary, choosing negative $\alpha$ mainly restricts the sampling of non reactive or regular trajectories within stable basins.

The distribution $\mathcal{P}^{\alpha}_{A}$ is approximated by a Markov chain of \emph{M} steps constructed by importance sampling, by means of the Metropolis algorithm. The sampling is done in the following way: at Markov chain step $m$, starting from the current path $\mathbf{z}^{m}$, a trial path $\tilde{\mathbf z}$ is generated with probability $P_{gen}$. Then, the trial path is accepted with a probability $P_{acc}$ and added to the Markov chain as $\mathbf{z}^{m+1}=\tilde{\mathbf z}$; otherwise, if the trial path is rejected, $\mathbf{z}^{m+1}=\mathbf {z}^m$. To ensure the convergence of the Markov chain towards the equilibrium distribution $\mathcal{P}_{A}$, we impose that the probability $\pi\left[\mathbf{z}\rightarrow\mathbf{z}'\right]$ to transit from a path $\mathbf{z}$ to a different path $\mathbf{z}'$ satisfies the detailed balance equation
\begin{equation}\label{bilandetaille}
\mathcal{P}^{\alpha}_A\left[\mathbf{z}\right]\pi\left[\mathbf{z}\rightarrow\mathbf{z}'\right]=\mathcal{P}^{\alpha}_A\left[\mathbf{z}'\right]\pi\left[\mathbf{z}'\rightarrow\mathbf{z}\right] 
\end{equation}
Taking account of the generating and acceptance probabilities $P_{gen}$ and $P_{acc}$, the transition probability $\pi$ reads
\begin{equation}\label{pi}
\pi\left[\mathbf{z}\rightarrow\mathbf{z}'\right]=\sum_{\tilde{\mathbf{z}}}P_{gen}\left[\mathbf{z}\rightarrow\tilde{\mathbf{z}}\right]\left\{\delta(\tilde{\mathbf{z}}-\mathbf{z}')P_{acc}\left[\mathbf{z}\rightarrow\tilde{\mathbf{z}}\right]+\delta(\tilde{\mathbf{z}}-\mathbf{z})(1-P_{acc}\left[\mathbf{z}\rightarrow\tilde{\mathbf{z}}\right])\right\} 
\end{equation}
where $\delta$ is the delta distribution and $\mathbf{z}'$ is either the old path $\mathbf{z}$ or the proposed path $\tilde{\mathbf{z}}$. 
The acceptance probability can be constructed from Eqs.~\eqref{bilandetaille} and ~\eqref{pi} as the Metropolis acceptance 
\begin{equation}
P_{acc}\left[\mathbf{z}\rightarrow\tilde{\mathbf{z}}\right]=\min\left\{ 1,\frac{\mathcal{P}^{\alpha}_A\left[\tilde{\mathbf{z}}\right]P_{gen}\left[\tilde{\mathbf{z}}\rightarrow\mathbf{z}\right]}{\mathcal{P}^{\alpha}_A\left[\mathbf{z}\right]P_{gen}\left[\mathbf{z}\rightarrow\tilde{\mathbf{z}}\right]}\right\}.\label{eq:11}\end{equation}
that is widely used in numerical simulations, as it has the main advantage of maximizing $P_{acc}$.

\subsubsection*{Shooting algorithm}

The standard shooting algorithm for deterministic dynamics is obtained by (i) selecting a state $\mathbf{x}_{t'}$ of the current trajectory, (ii) perturbing the momenta of each particle, and then generating from this selected state two segments, one backward of duration $t'$  and the other one forward of duration $\tau-t'$, in order to get a trial trajectory $\tilde{\mathbf{z}}$ of same duration, (iii) accepting or rejecting the new trajectory $\tilde{\mathbf{z}}$.  For deterministic dynamics, the total energy is constant. The perturbation step (ii) is done with the algorithm proposed by Stoltz \cite{stoltz2007path}, where trial momenta $\tilde{\mathbf{p}}$ are obtained from old momenta $\mathbf{p}$ as 
\begin{equation} \label{eq:stoltz}
 \tilde{\mathbf{p}}=\epsilon\mathbf{p}+\sqrt{1-\epsilon^{2}}\delta\mathbf{p}
\end{equation}
where $\epsilon$ is a tunable parameter and $\delta\mathbf{p}$ is drawn from a Gaussian
distribution of variance $k_BT$. 
The probability of having  trial momenta $\tilde{\mathbf{p}}$ from old momenta $\mathbf{p}$ with the Stoltz proposal for shooting~\cite{stoltz2007path} is written as
\begin{equation}\label{pbpp}
p\left(\mathbf{p}\rightarrow\tilde{\mathbf{p}}\right)=\left(\frac{1}{\sqrt{2\pi(1-\epsilon^{2})}}\right)^{3N}\exp\left\{-\frac{(\tilde{\mathbf{p}}-\epsilon\mathbf{p})^T(\tilde{\mathbf{p}}-\epsilon\mathbf{p})}{2(1-\epsilon^{2})}\right\}.
\end{equation} and ensures the detailed balance condition
\begin{equation}\label{eqht}
\frac{\exp\left\{-\beta \mathcal H(\tilde{\mathbf{x}}_{t'})\right\}p\left(\tilde{\mathbf{p}}_{t'}\rightarrow \mathbf{p}_{t'}\right)}{\exp\left\{-\beta \mathcal H(\mathbf{x}_{t'})\right\}p\left(\mathbf{p}_{t'}\rightarrow\tilde{\mathbf{p}}_{t'}\right)}=1
\end{equation} hence the probability flux between the current and perturbed momenta is balanced and the hamiltonian distribution is preserved.

The probability $p_{gen}[\mathbf{x}_{t'}\rightarrow \tilde{\mathbf{x}}_{t'}]$ of obtaining the shooting point $\tilde{\mathbf{x}}_{t'}$ for the trial trajectory from state $\mathbf{x}_{t'}$ selected in the current trajectory reads 
\begin{equation}\label{distrp}
p_{gen}[\mathbf{x}_{t'}\rightarrow \tilde{\mathbf{x}}_{t'}]=p\left(\mathbf{p}_{t'}\rightarrow\tilde{\mathbf{p}}_{t'}\right)
\end{equation} as only momenta are perturbed at the shooting point, while positions are left unchanged.

Inserting Eq.~\eqref{eq:9} in Eq.~\eqref{eq:11} and neglecting numerical approximations in the integration from $\tilde{\mathbf{x}}_t'$ to $\tilde{\mathbf{x}}_0$ (that indeed give in computations $\mathcal H(\tilde{\mathbf{x}}_t')\neq\mathcal H(\tilde{\mathbf{x}}_0)$) we have the Metropolis acceptance rule
 \begin{equation}\label{eq:p_ac2}
P_{acc}\left[\mathbf{z}\rightarrow\tilde{\mathbf{z}}\right]=\min\left\{ 1,\frac{\exp\left\{ \alpha L(\tilde{\mathbf{z}})-\beta \mathcal H(\tilde{\mathbf{x}}_0)\right\}\varphi_{A}^{\alpha}(\tilde{\mathbf{x}}_{0}) p_{gen}[\tilde{\mathbf{x}}_{t'}\rightarrow \mathbf{x}_{t'}] }{\exp\left\{ \alpha L(\mathbf{z})-\beta \mathcal H(\mathbf{x}_0)\right\}\varphi_{A}^{\alpha}(\mathbf{x}_{0})p_{gen}[\mathbf{x}_{t'}\rightarrow \tilde{\mathbf{x}}_{t'}] }\right\} \end{equation} where terms deriving from the forward and backward generation of the current and  trial path have cancelled because of the unit phase space compressibility of the Newtonian dynamics (see Ref.~\cite{dellago2002transition} for details).

Using in Eq.~\eqref{eq:p_ac2} the property of the Stoltz proposal (Eq.~\eqref{eqht}) and neglecting numerical approximations in the integration from $\tilde{\mathbf{x}}_t'$ to $\tilde{\mathbf{x}}_0$, Eq.~\eqref{eq:11} further simplifies in
\begin{equation} 
\label{eq:Pacc_shoot}
P_{acc}\left[\mathbf{z}\rightarrow\tilde{\mathbf{z}}\right]=\min \left\{ 1,\exp\left\{ \alpha L(\tilde{\mathbf{z}})-\alpha L(\mathbf{z})\right\}\left[\varphi_{A}^{\alpha}(\tilde{\mathbf{x}}_{0})/\varphi_{A}^{\alpha}(\mathbf{x}_{0})\right] \right\}.
\end{equation}

\subsubsection*{Shifting algorithm}\label{waste-recycling}

The second Monte Carlo move in trajectory space is based on the shifting algorithm, supplemented with a waste-recycling estimator.~\cite{adjanor2011waste} $\mathcal N$ trial trajectories $\tilde{\mathbf{z}}_j$ are constructed from $\mathbf z$ as follows: the duration of the trajectory $\mathbf {z}$ is doubled selecting a random duration $\nu\delta t$ and integrating two segments backward and forward, starting from $\mathbf{x}_{0}$ and $\mathbf{x}_{\tau}$, along $\nu$ and $\mathcal{N}-\nu$ time steps, respectively. Adding these segments to the current trajectory, one obtains a ``buffer'' trajectory $\mathbf{\zeta}=\left\{\mathbf{x}_{n \delta t}\right\}_{-\nu\leq n\leq2\mathcal{N}-\nu}$ of total duration $2\mathcal{N}\delta t$, containing $\mathcal{N}$ possible trial paths. The conditional probability of obtaining the ``buffer'' trajectory $\mathbf{\zeta}$ starting from the current trajectory $\mathbf{z}$ is indicated as  $P_{cond}(\mathbf{\zeta}|\mathbf{z})$. The probability weight of each trial path $\tilde{\mathbf{z}}_j$ is then\begin{equation}\label{eq:19}
\mathcal{P}^{\alpha}_{A}(\tilde{\mathbf{z}}_{j})=\frac{1}{Z^{\alpha}_A}\exp\left\{ \alpha L(\tilde{\mathbf{z}}_{j})-\beta \mathrm H(\tilde{\mathbf{z}}_{j})\right\}\varphi_{A}^{\alpha}(\tilde{\mathbf{x}}_{0,j}) 
\end{equation}
if a constraining function $\varphi_{A}^{\alpha}(\tilde{\mathbf{x}}_{0,j})$ linking the initial state $\tilde{\mathbf{x}}_{0,j}$ of each trial path $\tilde{\mathbf{z}}_{j}$ trajectories to state A is used, as in Eq.~\eqref{eq:9}. Index \emph{j} runs over the \emph{$\mathcal{N}$} possible paths on the ``buffer'' trajectory.

We introduce an action
\begin{equation}
-s_{\alpha,j}=\alpha L(\tilde{\mathbf{z}}_{j})-\beta \mathrm H(\tilde{\mathbf{z}}_{j}) 
\end{equation}
so as to rewrite the trajectory probability in the biased ensemble (Eq.~\eqref{eq:19}) as 
\begin{equation}\mathcal{P}^{\alpha}_{A}(\mathbf{z}_{j})=\frac{1}{Z^{\alpha}_A}\varphi_{A}^{\alpha}(\tilde{\mathbf{x}}_{0,j})\exp\left\{-s_{\alpha,j}\right\}. \end{equation} We now define the selecting probability ${P}_{sel}(\tilde{\mathbf{z}}_{j}|\mathbf{\zeta})$ of selecting a trial trajectory $\tilde{\mathbf{z}}_{j}$ from the buffer trajectory $\mathbf{\zeta}$ as
\begin{equation}\label{pisel}
{P}_{sel}(\tilde{\mathbf{z}}_{j}|\mathbf{\zeta})=\frac{\varphi_{A}^{\alpha}(\tilde{\mathbf{x}}_{0,j})\exp\left\{-s_{\alpha,j}\right\}}{\mathcal{R}_{\alpha}}
\end{equation}
where we introduced the Rosenbluth factor 
\begin{equation}\label{pmarg}
\mathcal{R}_{\alpha}=\sum_{j=1}^{\mathcal N}\varphi_{A}^{\alpha}(\tilde{\mathbf{x}}_{j})\exp\left[-s_{\alpha,j}\right].
\end{equation}
Resorting to Bayes theorem, $P_{sel}$ can be written as the posterior likelihood probability~\cite{AM2010} of having $\tilde{\mathbf{z}}_{j}$ given the ``buffer'' trajectory $\mathbf{\zeta}$:
\begin{equation}\label{bayes}
P_{sel}(\tilde{\mathbf{z}}_{j}|\mathbf{\zeta})=\frac{P_{cond}(\mathbf{\zeta}|\tilde{\mathbf{z}}_{j})\mathcal{P}^{\alpha}_{A}(\tilde{\mathbf{z}}_{j})}{P_{marg}^{\alpha}(\mathbf{\zeta})}
\end{equation}
where $P_{cond}(\mathbf{\zeta}|\tilde{\mathbf{z}}_{j})$ is the conditional probability of constructing a ``buffer'' path $\mathbf{\zeta}$ from the trial path, and because of the deterministic dynamics $P_{cond}(\mathbf{\zeta}|\tilde{\mathbf{z}}_j)=1/\mathcal{N}$. $P_{marg}^{\alpha}(\mathbf{\zeta})$ is the marginal probability associated with the buffer path: comparing Eq.~\eqref{bayes} with Eq.~\eqref{pisel}, we see that
\begin{equation}\label{pmarg3}
P_{marg}^{\alpha}(\mathbf{\zeta})=\mathcal{R}_{\alpha}\frac{1}{Z^{\alpha}_A}\frac{1}{\mathcal N}.
\end{equation}

Let us now consider a Monte Carlo move between two paths both contained in the buffer trajectory $\zeta$ , i.e. from the current path $\mathbf{z}$ to $\tilde{\mathbf{z}}_{j}$, whose associated transition probability $\pi[\mathbf{z}\rightarrow \tilde{\mathbf{z}}_{j}]$, as in Eq.~\eqref{bilandetaille}, obeys a detailed balance with respect to the prior distribution $\mathcal{P}^{\alpha}_{A}$:
\begin{equation}\label{bildetwr}
\mathcal{P}_A^{\alpha}\left[\mathbf{z}\right]\pi\left[\mathbf{z}\rightarrow\tilde{\mathbf{z}}_{j}\right]=\mathcal{P}^{\alpha}_A\left[\tilde{\mathbf{z}}_{j}\right]\pi\left[\tilde{\mathbf{z}}_{j}\rightarrow\mathbf{z}\right].
\end{equation} Defining the transition probability
\begin{equation}\label{bildetpost}
\pi\left[\mathbf{z}\rightarrow\tilde{\mathbf{z}}_{j}\right]=P_{sel}(\tilde{\mathbf{z}}_{j}|\mathbf{\zeta})P_{cond}(\mathbf{\zeta}|{\mathbf{z}})
\end{equation} the detailed balance in Eq.~\eqref{bildetwr} can be recasted as
\begin{equation}\label{bildetshift}
P_{sel}(\tilde{\mathbf{z}}_{j}|\mathbf{\zeta})P_{cond}(\mathbf{\zeta}|\mathbf{z})\mathcal{P}^{\alpha}_{A}(\mathbf{z})={P}_{sel}(\tilde{\mathbf{z}}_{\nu}|\mathbf{\zeta})P_{cond}(\mathbf{\zeta}|\tilde{\mathbf{z}}_j)\mathcal{P}^{\alpha}_{A}(\tilde{\mathbf{z}}_{j})
\end{equation}
where $\mathbf{z}=\tilde{\mathbf{z}}_{\nu}$ and ${P}_{sel}(\tilde{\mathbf{z}}_{\nu}|\mathbf{\zeta})$ is the probability to transit from $\tilde{\mathbf{z}}_{j}$ to $\mathbf{z}$. As a consequence, the shifting procedure leaves the probability distribution $\mathcal{P}^{\alpha}_{A}$ invariant. Moreover, recalling that the deterministic dynamics entails $P_{cond}(\mathbf{\zeta}|\mathbf{z})=P_{cond}(\mathbf{\zeta}|\tilde{\mathbf{z}}_j)$, the detailed balance in Eq.~\eqref{bildetshift} further simplifies into 
\begin{equation}\label{bildetshift2}
P_{sel}(\tilde{\mathbf{z}}_{j}|\mathbf{\zeta})\mathcal{P}^{\alpha}_{A}(\mathbf{z})={P}_{sel}(\tilde{\mathbf{z}}_{\nu}|\mathbf{\zeta})\mathcal{P}^{\alpha}_{A}(\tilde{\mathbf{z}}_{j}).
\end{equation}

We conclude this section by extending the detailed balance of Eq.~\eqref{bildetwr} for trajectories $\mathbf{z}$ and $\tilde{\mathbf{z}}_{j}$ belonging to two \textit{different} buffer paths $\zeta$ and $\tilde{\zeta}$ respectively. Writing as in Eq.~\eqref{bayes} expressions for $P_{sel}(\mathbf{z}|\mathbf{\zeta})$ and $P_{sel}(\tilde{\mathbf{z}}_{j}|\tilde{\mathbf{\zeta}})$, and using these results in Eq.~\eqref{bildetwr} we obtain
\begin{equation}\label{bildettot}
P_{marg}^{\alpha}(\zeta)P_{sel}(\mathbf{z}|\mathbf{\zeta})\pi\left[\mathbf{z}\rightarrow\tilde{\mathbf{z}}_{j}\right]=P_{marg}^{\alpha}(\tilde{\zeta})P_{sel}(\tilde{\mathbf{z}}_{j}|\tilde{\mathbf{\zeta}})\pi\left[\tilde{\mathbf{z}}_{j}\rightarrow\mathbf{z}\right] 
\end{equation} Eq.~\eqref{bildettot} is indeed a detailed balance condition for the alternate shooting and shifting moves, and samples the buffer trajectories $\zeta$. This implies that the distribution $P_{marg}^{\alpha}$ is invariant along the sampling algorithm. $P_{marg}^{\alpha}$ is therefore suitable to be used as an input path probability weight required by the unbiasing algorithm MBAR, that will be presented and used in the following Sections.

\section{Reaction-rate constants calculation \label{sec:Reaction-rate-constants-calculation}}

\subsection{Rate constants theory} 

Here we briefly recall how reaction rates are computed in the TPS framework. The reactivity of the sampled paths is recovered by means of a time correlation function with respect to initial and final states of the paths: a trajectory is said to be reactive if it starts in the reactants A basin and ends in the products B basin. The time correlation function~\cite{dellago2002transition,W2003} is 
\begin{equation}
\label{eq:correl}
C(t)=\frac{\left\langle h_{A}(\mathbf{x}_{0})h_{B}(\mathbf{x}_{t})\right\rangle }{\left\langle h_{A}(\mathbf{x}_{0})\right\rangle }\end{equation}
where brackets $\left\langle \cdot\right\rangle $ indicate averages taken over the equilibrium trajectory ensemble and the indicator function $h_{\Omega}$ is defined as in Eq.~\eqref{ha}. 
$C(t)$ can be understood as the conditional probability of observing a trajectory of duration $t$ ending in state B, knowing that  it started in state A~\cite{chandler1978statistical,dellago2002transition,W2003}, 
and approaches its asymptotic value exponentially  as
\begin{equation}
\label{eq:correl_exp}
C(t)\approx \rho_{B}^{eq}\left(1-\exp\left\{ -t/\tau_{rxn}\right\} \right)
\end{equation}
where $\rho_{B}^{eq}$ is the equilibrium occupation probability of state B, and the parameter $\tau_{rxn}\equiv\left(k_{A\rightarrow B}+k_{B\rightarrow A}\right)^{-1}$ is the characteristic reaction time of the system, given by the forward and backward reaction constants $k_{A\rightarrow B}$ and $k_{B\rightarrow A}$, respectively.

Note that the basic assumption required to compute reaction rate constants of rare events from the correlation function $C(t)$ is the presence of a well separated time scale for processes occurring between 'fast' intra-funnel relaxation, having a typical time constant $\tau_{mol}$, and activated processes indicating passages \emph{between} funnels, needing a much longer time scale, of the order of $\tau_{rxn}$ ~\cite{chandler1978statistical}. 

For times in the intermediate time regime $\tau_{mol}<t\ll\tau_{rxn}$, the correlation function in Eq.~\eqref{eq:correl_exp} can indeed be expressed by its first order expansion 
\begin{equation} \label{eq:c_kt}
C(t)\approx k_{A\rightarrow B}t 
\end{equation}
where the detailed balance condition $k_{A\rightarrow B}\rho_{A}^{eq}=k_{B\rightarrow A}\rho_{B}^{eq}$ has been used to eliminate $\rho_{B}^{eq}$.  Hence, the slope of $C(t)$ for this intermediate time regime gives direct access to reaction rates. The reactive probability flux flowing from state A towards B per unit time, defined by $k(t)\equiv\frac{dC(t)}{dt}$, displays a plateau corresponding to the forward phenomenological reaction constant $k_{A\rightarrow B}$.~\cite{chandler1987introduction}

Let us point out that these results, obtained with a macroscopic 'flux over population' probability approach, can be recovered by a Bennett-Chandler formalism ~\cite{chandler1987introduction}, based on microscopic quantities (positions and momenta, see Ref.~\cite{chandler1978statistical}). Finally, we recall that the phenomenological forward rate constant $k_{A\rightarrow B}$ computed from Eq.~\eqref{eq:c_kt} is related to the rate $k_{TST}$ given by the transition state theory (TST)~\cite{HTB1990, chandler1987introduction} 
\begin{equation}\label{eq:react_flux}
k(t)=\kappa(t) k_{TST}
\end{equation}
where $\kappa(t)$ is the \textit{transmission factor}, and $k_{TST}$ reads 
\begin {equation}\label{kappa_tst}
 k_{TST}=\left(\frac{k_{B}T}{h}\right)\exp\left(-\beta\Delta F_{A\rightarrow B}\right)
\end {equation}
with $\Delta F_{A\rightarrow B}$ the height of the free energy barrier separating basins A and B, $h$ the Planck constant and $\beta =\frac{1}{k_{B}T}$ the inverse temperature.  The transmission factor $\kappa$ is always lower than one, and is introduced to take into account trajectories started in basin A that reach the saddle point but fall back to state A, instead of ending in state B: these occurrences are called recrossings events. The transmission coefficient usually reaches a steady value, depending on the temperature and the reaction coordinate chosen to localize the barrier. At this plateau value of $\bar\kappa<1$, and the reactive flux corresponds to $k_{A\rightarrow B}=\bar\kappa k_{TST}$, always lower than the TST value.

\subsection{Rate constants with biased sampling and \emph{waste-recycling}}\label{rarecostbiased}

In numerical experiments, the computation of reaction constants by direct evaluation of $C(t)$ at times longer than the intrafunnel relaxation time $\tau_{mol}$ means performing very long molecular dynamics trajectories. To estimate $C(t)$ in the TPS framework computing relatively short trajectories, one resorts to a factorization of the correlation function in (i) a static quantity related to $k_{TST}$, thus dependent on the free-energy barrier and calculated through an umbrella-sampling technique, and (ii) a dynamic factor related to $\kappa(t)$ given by the time derivative of the probability of reaching basin B at times shorter than the whole trajectory length (see for instance Refs. ~\cite{dellago1999calculation,DBCD1998})

We propose herein a variant strategy: reaction constants will be calculated directly by averaging indicator functions on short trajectories, as in Eq.~\eqref{eq:correl}, once the fraction of reactive paths is significantly enhanced by introducing an appropriate bias favoring reactions between basins A and B. 

The correlation function in Eq.~\eqref{eq:correl} can be intended as an average over all performed trajectories - i.e. over all successive steps \emph{m} of the Markov chain - of the \emph{reactivity} $\mathcal{A}(\mathbf{z}^{m})$, defined as 
\begin{equation}\label{eq:a_react}
 \mathcal{A}(\mathbf{z}^{m})= h_{A}(\mathbf{x}_{0}^{m})h_{B}(\mathbf{x}_{\tau}^{m})
\end{equation}
For an unbiased TPS simulation, the probability $\mathcal{P}^{\alpha}_{A}$ of Eq.~\eqref{eq:9} with $\alpha=0$ is sampled: we have $C(t)=\left\langle \mathcal{A}\right\rangle_0$, where brackets $\left\langle \cdot\right\rangle_0 $ correspond to averages over a canonical trajectory ensemble, and the trajectory distribution of Eq.~\eqref{eq:9} ensures $\left\langle h_{A}(\mathbf{x}_{0})\right\rangle_0 =1$ because of $\varphi_{A}^{\alpha}$.

In a context of biased TPS, averages on Markov chains are taken on the biased trajectory distribution, hence we denote biased averages of the correlation function as $C_{\alpha}(t)=\left\langle \mathcal{A}\right\rangle _{\alpha}$, where index $\alpha$ accounts for the current bias. Herein, index $\alpha$ will indicate all observables obtained from a biased distribution, where $\alpha=0$ stands for the equilibrium canonical ensemble average.
We estimate the correlation function $C_{\alpha}(t)= \left\langle \mathcal{A}\right\rangle _{\alpha}$ using an estimator denoted by $ \mathbb{I}_{\alpha}^M$, which consists in taking the average over the Markov chain of lenght $M$ as
\begin{equation}\label{est_corrf}
\mathbb{I}_{\alpha}^M\left[\mathcal A\right]=\frac{1}{M}\sum_{m=1}^{M}\mathcal{A}^{m}_{\alpha}
\end{equation}
$\mathcal{A}(\mathbf{z}^{m}_{\alpha})=\mathcal{A}^{m}_{\alpha}$ denotes a reactivity value coming from a biased trajectory $\mathbf{z}_{\alpha}$ distributed according to $\mathcal{P}^{\alpha}_{A}$. 
Estimates given by $\mathbb{I}_{\alpha}^M$ are however not optimal~\cite{wr1}.

A \emph{waste-recycling} (WR) estimator ~\cite{wr1,wr2} is associated to the multiple proposal sampler for the shifting move of Sec.~\ref{waste-recycling} to obtain more accurate estimates. Waste recycling consists in including information about all possible proposals contained in the buffer trajectory $\zeta$. The reactivity at Markov chain step \emph{m} in a given ensemble $\alpha$ has to be first averaged over the $\mathcal{N}$ trajectories contained in the buffer path $\zeta^{m}_{\alpha}$ as 
\begin{equation}\label{reactwr}
\overline{\mathcal{A}}^{m}_{\alpha}=\sum_{j=1}^{\mathcal N}\mathcal{A}^{m}_{\alpha,j}\frac{\varphi_{A}^{\alpha}(\mathbf{x}^m_{0,j})\exp\left[-s^{m}_{\alpha,j}\right] }{\mathcal{R}^{m}_{\alpha}}=\sum_{j=1}^{\mathcal N}\mathcal{A}^{m}_{\alpha,j}\varphi_{A}^{\alpha}(\mathbf{x}^m_{0,j})\exp\left[-s^{m}_{\alpha,j}+S^{m}_{\alpha}\right]
\end{equation}
where we write $\overline{\mathcal{A}}^{m}_{\alpha}=\overline{\mathcal{A}}(\mathbf{\zeta}^{m}_{\alpha})$ and define the Rosenbluth factor
\begin{equation}\label{pmarg2}
\mathcal{R}^{m}_{\alpha}\equiv \exp{\left[-S^{m}_{\alpha}\right]}
\end{equation} as proportional to the marginal probability $P^{\alpha}_{marg}(\mathbf{\zeta}^{m}_{\alpha})$ of Eq.~\eqref{pmarg}, associated to the ``buffer'' trajectory $\mathbf{\zeta}^{m}_{\alpha}$ corresponding to Markov chain step $m$.  Correlation functions for reactive paths are therefore estimated in a way similar to Eq.~\eqref{est_corrf}, with the WR estimator
\begin{equation}\label{est_corrf_wr1}
\mathbb{J}^{M}_{\alpha}\left[\mathcal A\right]=\frac{\sum_{m=1}^{M}\sum_{j=1}^{\mathcal N}\mathcal{A}^{m}_{\alpha,j}\varphi_{A}^{\alpha}(\mathbf{x}^m_{0,j})\exp\left[ -s^{m}_{\alpha,j}+S^{m}_{\alpha}\right]}{\sum_{m=1}^{M}\sum_{j=1}^{\mathcal N}\varphi_{A}^{\alpha}(\mathbf{x}^m_{0,j})\exp\left[ -s^{m}_{\alpha,j}+S^{m}_{\alpha}\right]}=\frac{1}{M}\sum_{n=1}^{M}\overline{\mathcal{A}}^{m}_{\alpha}
\end{equation} and again $C_{\alpha}(t)\approx \mathbb{J}_{\alpha}\left[\mathcal A\right]$.
The calculation of rate constants $k_{A\rightarrow B}^{\alpha}$ for the given $\alpha$-ensemble follows from Eq.~\eqref{eq:c_kt}.

\subsection{Unbiasing rate constants: the MBAR algorithm}\label{MBAR}

A suitable unbiasing algorithm is needed in order to recover canonical ensemble correlation functions from reactivity values witnessed in a Lyapunov biased path ensemble.  The canonical equilibrium values of $C_0(t)$ can in principle be obtained by estimating reactivities $\mathcal{A}^{m}_{\alpha}$, computed in any $\alpha$-biased path ensemble, resorting to an adequate unbiasing algorithm. We define an unbiasing WR estimator $\mathbb{J}_{\theta,\alpha}$, where the left subscript $\theta$ indicates the ensembles in which we are interested in measuring averages, while the right subscript $\alpha$ refers to the ensemble that our Lyapunov biased TPS will effectively sample. Equilibrium values for $C_0(t)$ ($\theta=0$) are retrieved from reactivities computed in any $\alpha$-biased ensemble as
\begin{equation}\label{est_corrf_wr_unb}
C_0(t)\approx\mathbb{J}^{M}_{0,\alpha}\left[\mathcal{A}\right]=\frac{\sum_{n=1}^{M}\sum_{j=1}^{\mathcal N}\mathcal{A}^{m}_{\alpha}\varphi_{A}^{\alpha}(\mathbf{x}^m_{0,j}){\varphi_{A}^{\alpha}}^{-1}(\mathbf{x}^m_{0,j})\exp\left[ -s^{m}_{\alpha,j}+S^m_{\alpha}+s^{m}_{\alpha,j}\right] 
}{\sum_{n=1}^{M}\sum_{j=1}^{\mathcal N}\varphi_{A}^{\alpha}(\mathbf{x}^m_{0,j}){\varphi_{A}^{\alpha}}^{-1}(\mathbf{x}^m_{0,j})\exp\left[ -s^{m}_{\alpha,j}+S^m_{\alpha}+s^{m}_{\alpha,j}\right]}
\end{equation}
Unbiasing the sampling consists of correcting for the bias $\varphi_{A}^{\alpha}(\mathbf{x}^m_{j})\exp\left[-s^{m}_{\alpha,j}\right]$. However, the variance associated to the unbiasing estimator in Eq.~\eqref{est_corrf_wr_unb} would be too large to consider estimates reliable~\cite{shirts2008statistically}. This well-known fact results from the lack of overlap between the sampled and measured distributions. We therefore carry out a series of simulations for a set of $\alpha$ values ranging from 0 to a maximum value $\alpha_{max}$ so to ensure overlap between successive sampled distributions.

Our choice is then to use the multistate Bennett acceptance ratio (MBAR) instead of the estimator of Eq.~\eqref{est_corrf_wr_unb}. MBAR is a method elaborated by Shirt and Chodera \cite{shirts2008statistically,minh2009optimal}, which aims at minimizing the statistical variance associated to the estimates. Following these authors, we briefly expose the principles of this procedure in a context of biased path ensembles.

To use the MBAR method in the waste-recycling framework of Sec.~\ref{rarecostbiased}, we take as probability weights corresponding to a given $\alpha$ ensemble at each Markov chain step $m$ the marginal probability $P_{marg}^{\alpha}(\zeta^m_{\alpha})$ of Eq.~\eqref{pmarg2}. This is possible because, thanks to the detailed balance of Eq.~\eqref{bildettot}, $P_{marg}^{\alpha}$ is preserved. 

Once one knows weights $\exp{\left[-S_{\alpha}\right]}$ related (via Eqs.~\eqref{pmarg2} and \eqref{pmarg3}) to $P_{marg}^{\alpha}(\zeta_{\alpha})$ for each $\alpha$-ensemble, reactivity averages $\left\langle\mathcal{A}\right\rangle _{\alpha'}$  for every ensemble $\alpha'\neq \alpha$ can be computed resorting to the importance sampling identity 
\begin{equation}
\frac{\left\langle \mathcal{A}\exp{\left[-S_{\alpha'}\right]}\right\rangle _{\alpha}}{\left\langle \mathcal{A}\exp{\left[-S_{\alpha}\right]}\right\rangle _{\alpha'}}=\frac{Z_{\alpha'}}{Z_{\alpha}}\label{eq:10}
\end{equation}
where we used the partition functions ${Z}_{\alpha}=\int \mathcal D\zeta \exp{\left[-S_{\alpha}(\zeta)\right]} $.
For a set of \emph{K} different values of the bias $\alpha$, a set (namely, a Markov chain) of $\mathcal{M_{\alpha}}$ buffer trajectories are sampled for each bias value. An estimate of $C_{\alpha}(t)$ is given by the MBAR estimator $\mathbb{K}$ for the waste-recycling averaged reactivity $\overline{\mathcal A}_{\alpha}$ of Eq.~\eqref{reactwr} as
\begin{equation}\label{estim}
\mathbb{K}^{\mathcal{M}_{\alpha}}_{\alpha}\left[\mathcal A \right]=\sum_{m=1}^{\mathcal{M}_{\alpha}} W_{m,\alpha} \overline{\mathcal A}^m_{\alpha}
\end{equation} where weights $W_{m,\alpha}$ are given by the expression 
\begin{equation}\label{matrW2}
W_{m,\alpha}=\hat{Z}_{\alpha}^{-1}\frac{\exp{\left[-S_{\alpha}^m\right]}}{\sum_{k=1}^{K}\mathcal{M}_{k}\hat{Z}_{k}^{-1}\exp{\left[-S_{\alpha}^m\right]}}.
\end{equation} and $\hat{Z}_{\alpha}$ are estimators for the partition functions ${Z}_{\alpha}$ with minimal asymptotic covariance (see Eq.~\eqref{eq.estimating} and Ref.~\cite{minh2009optimal}). Note that the denominator in Eq.~\eqref{matrW2} indicates that each weight $W_{m,\alpha}$ takes into account contributions from all other ensembles $k=0,...,\alpha,...,K.$

Introducing also partition functions $Z_{\mathcal A_{\alpha}}\equiv\int\mathcal{D}\mathbf{\zeta}\mathcal A \exp{\left[-S_{\alpha}(\zeta)\right]}$, the uncertainty can be estimated as
\begin{equation}\label{estim}
\rm var(\mathbb{K}_{\alpha}[{\mathcal A}])\approx{\mathbb{K}_{\alpha}[\mathcal A]}^2 (\rm{var}(\hat Z_{\mathcal A_{\alpha}})+ \rm {var}(\hat{Z}_{\alpha})-2 \rm {cov}(\hat Z_{\mathcal A_{\alpha}},\hat{Z}_{\alpha})).
\end{equation}

Canonical equilibrium values of the correlation function $C_0(t)$ and corresponding values of the reaction rate constants can be recovered once we consider estimates in Eq.~\eqref{estim} with $\alpha=0$. We emphasize that this is by now the first application of the MBAR unbiasing method based on marginal probabilities, able to estimate observables computed with a path sampling algorithm supplemented by waste-recycling. Numerical recipes to obtain these estimations have been furnished by J. Chodera~\cite{chodera_site}.

\section{Numerical results}\label{Numerical results}

\subsection{LJ$_{38}$}\label{lj38}

In order to test the Lyapunov biased TPS algorithm (LyTPS), we first consider a Lennard-Jones 38 cluster. This well-known benchmark system is often uysed to assess the efficiency of sampling algorithms. It and has been widely explored in literature for its rich thermodynamic properties: its potential energy landscape presents two main basins: a deep and narrow funnel containing the global energy minimum, a face-centered cubic truncated octahedron structure (FCC), and a separate, wider, funnel leading to a large number of icosahedral structures (ICO) of slightly higher energies. Although the configuration with the lowest potential energy corresponds to the FCC one, the greater configurational entropy associated with a large number of local minima in the icosahedral funnel make this second configuration much more stable at higher temperatures. As temperature increases, LJ$_{38}$ undergoes several structural transitions. First, a solid-solid transition occurs at $T_{ss}=0.12 \frac{\epsilon}{k_{B}}$ when the octahedral FCC structure gives place to the icosahedral ones. Secondly, above $T_{sl}=0.18 \frac{\epsilon}{k_{B}}$, the outer layer of the cluster melts, while the core remains of icosahedral structure.~\cite{MF2006} 

The Lennard-Jones potential is given by the usual expression
\begin{equation}
V\left(\left\{ \mathbf{q}^{j}\right\}_{j=1,...,N}\right)=4\epsilon\sum_{j<k}
  \left[\left(\frac{\sigma}{r_{jk}}\right)^{12}-\left(\frac{\sigma}{r_{jk}}\right)^{6}\right]
\end{equation}
where $\mathbf{q}^{j}=(q^j_{x},q^j_{y},q^j_{z})$ is the position of
the $j$-th atom, $r_{jk}=\left|\mathbf{q}^{j}-\mathbf{q}^{k}\right|$
is the distance between atoms $j$ and $k$, $\epsilon$ is the pair well
depth and $2^{1/6} \sigma$ is the equilibrium pair separation. In
addition, a trapping potential prevents the evaporation of clusters particles at finite temperature: if the distance between the position ${\bf q}$ of a particle and the center of the trap ${\bf q_c}$ exceeds $2.25 \sigma$, this  particle feels a potential
$|{\bf q}-{\bf q_c}|^3$.

The bond-orientational order parameters $Q_{l}$ defined as~\cite{SNR1981,SNR1983} 
\begin{equation}
Q_{l}=\left(\frac{4\pi}{2l+1}\frac{1}{{{N}}_{b}^{2}}\sum_{m=-l}^{l}\left|\sum_{N_b}Y_{lm}\left(\theta_{jk},\phi_{jk}\right)\right|^{2}\right)^{1/2}\;
\end{equation}
where the $Y_{lm}(\theta,\phi)$ are spherical harmonics and $\theta_{jk}, \phi_{jk}$ are the polar and azimuthal angles of a vectors corresponding to bonds between the $N_b$ pairs of atoms. The $Q_l$ are traditionally used to identify different crystalline symmetries. The parameter $Q_{4}$ is often used to distinguish between the icosahedral and cubic structures, for which it has values around 0.02 and 0.18 respectively.~\cite{NCFD2000} 

Monte Carlo sampling fails to equilibrate the two funnels, and global optimization methods are unable to find its global energy minimum~\cite{W2003}. Hence, several elaborated algorithm have been employed in the past to study the thermodynamic equilibrium of this system, such as parallel tempering,~\cite{NCFD2000,CNFD2000,MF2006} basin-sampling techniques,~\cite{BWC2006} Wang-Landau approaches~\cite{PCABD2006} or path-sampling methods.~\cite{AAC2006,MP2007,AM2010}

Standard transition path sampling~\cite{MP2007} and discrete path sampling~\cite{W2002} (DPS) have been already used to study transitions between the two funnels of LJ$_{38}$. However, in the case of TPS the large number of metastable states separating the two main
basins prevented the traditional shooting and shifting algorithm to identify reactive paths, despite previous success for smaller LJ clusters.~\cite{DBC1998} Authors had to resort to a two-ended approach linking the two minima to find trajectories with the same energy of those found by DPS approach.~\cite{MP2007} The main drawbacks of this TPS method were a lack of ergodicity and a very large computational cost.

Conversely, DPS has been more succesfull in this task.  This method uses eigenvector following and graph transformation ~\cite{TW2006} to compute the overall transition rate between two regions of phase space. To the best of our knowledge, this is by now the most successful approach to computing reaction rates in LJ$_{38}$.~\cite{TW2006} In particular, reaction rate constants for transitions between the two solid structures have been computed using DPS~\cite{W2002,W2004,BW2006} at different temperatures.

We use here the Lyapunov biased TPS algorithm to investigate structural transitions in LJ$_{38}$ for temperatures above and below the solid-solid transition temperature $T_{ss}=0.12 $, spanning a temperature range from 
$T=0.10$ to $T=0.15$. Our simulations required about $10^2$ hours of cpu time to observe reactive trajectories between the two main funnels. Reaction constants, computed with the method exposed in Sec.~\ref{sec:Reaction-rate-constants-calculation}, can be compared to values obtained with the discrete path sampling approach.~\cite{MP2007}

\subsubsection{FCC-ICO reactive paths}\label{LJreactivepaths}

In order to thermalize the system at a given temperature, Langevin dynamics are run for $1000$ time steps, using a friction parameter $\gamma \delta t =1$. The last configuration is used as the starting point of the first deterministic trajectory of the TPS simulations. At each new temperature, a new preliminary Langevin dynamics is performed. 
 
In simulations, trajectories consist of $\mathcal N=700$ time steps, each step of duration $\delta t=10^{-2}$. Deterministic trajectories are obtained with the Verlet algorithm~\cite{stoltz2007path}, and then selected following the Lyapunov biased TPS algorithm described in Sec.~\ref{sec:Transition-path-sampling}. For each temperature, 25 different biased path distributions are sampled, for values of the control parameter $\alpha$ ranging linearly from $\alpha=1$ to $\alpha=2500$, in order to obtain reactive paths for $\alpha=2500$ and ensure a sufficient overlap between distributions sampled for different $\alpha$ values. The unbiased distribution corresponding to $\alpha=0$ has been simulated with TPS as well. A Markov sequence of $5000$ biased TPS shooting and shifting moves is performed, in order to ensure an ergodic sampling.

\begin{figure}
  \begin{center}
\includegraphics{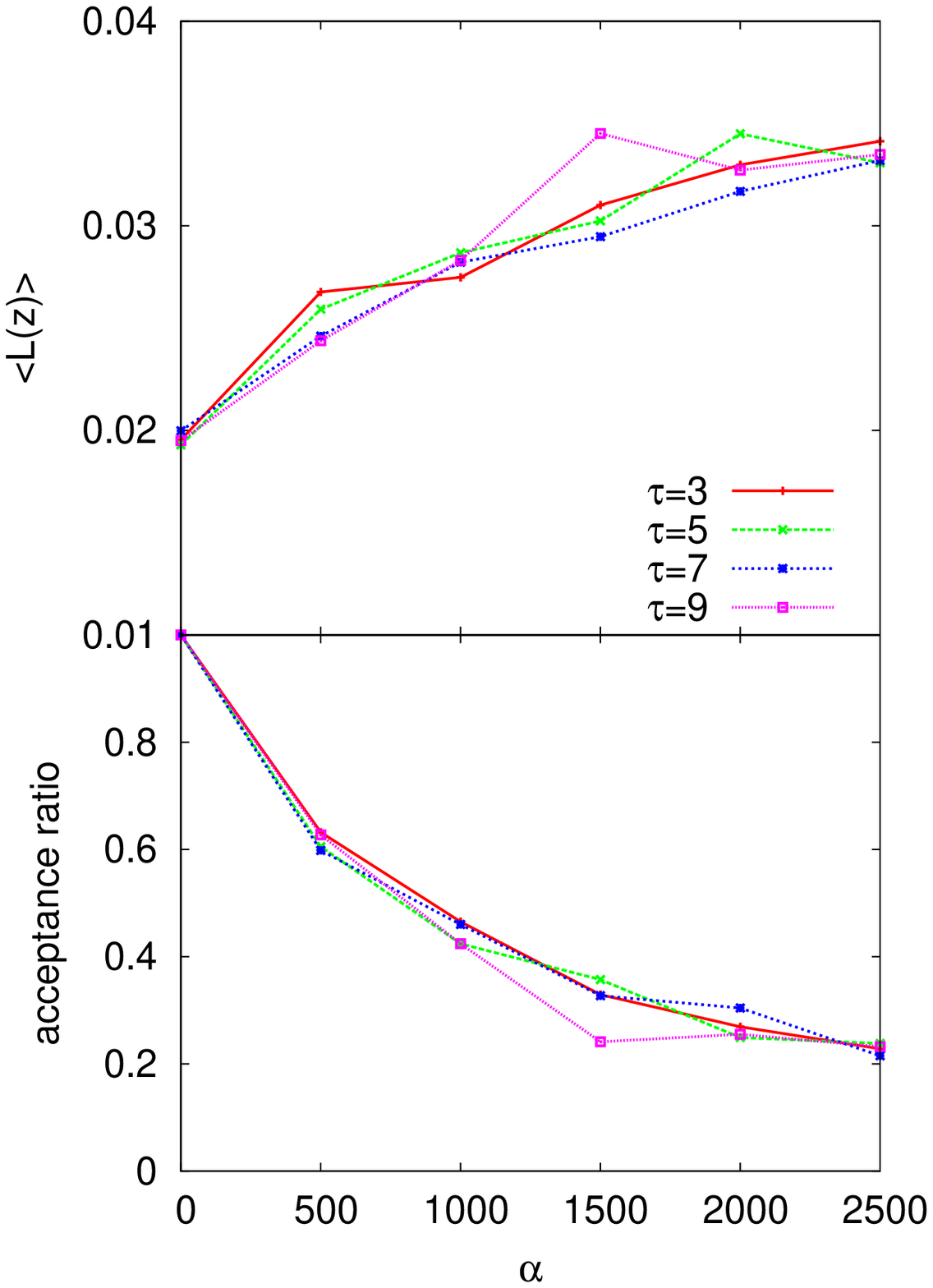} 
  \end{center}
\caption{{\bf Top}: Average value of the Lyapunov indicator $L(\mathbf z)$ over a Markov chain of $1000$ trajectories, starting from the FCC basin at $T=0.12$, as a function of the control parameter $\alpha$ used in the simulations, for different trajectory lengths $\tau$. Increasing $\alpha$ increases the mean Lyapunov indicator and enables trajectories to explore barriers and transition states.  Average Lyapunov indicators are almost independent of the trajectory length $\tau$. {\bf Bottom}: Acceptance ratio, given by Eq.~\eqref{eq:Pacc_shoot} for the same simulations.}
  \label{fig:acceptance}
\end{figure}

\begin{figure}
  \begin{center}
\includegraphics{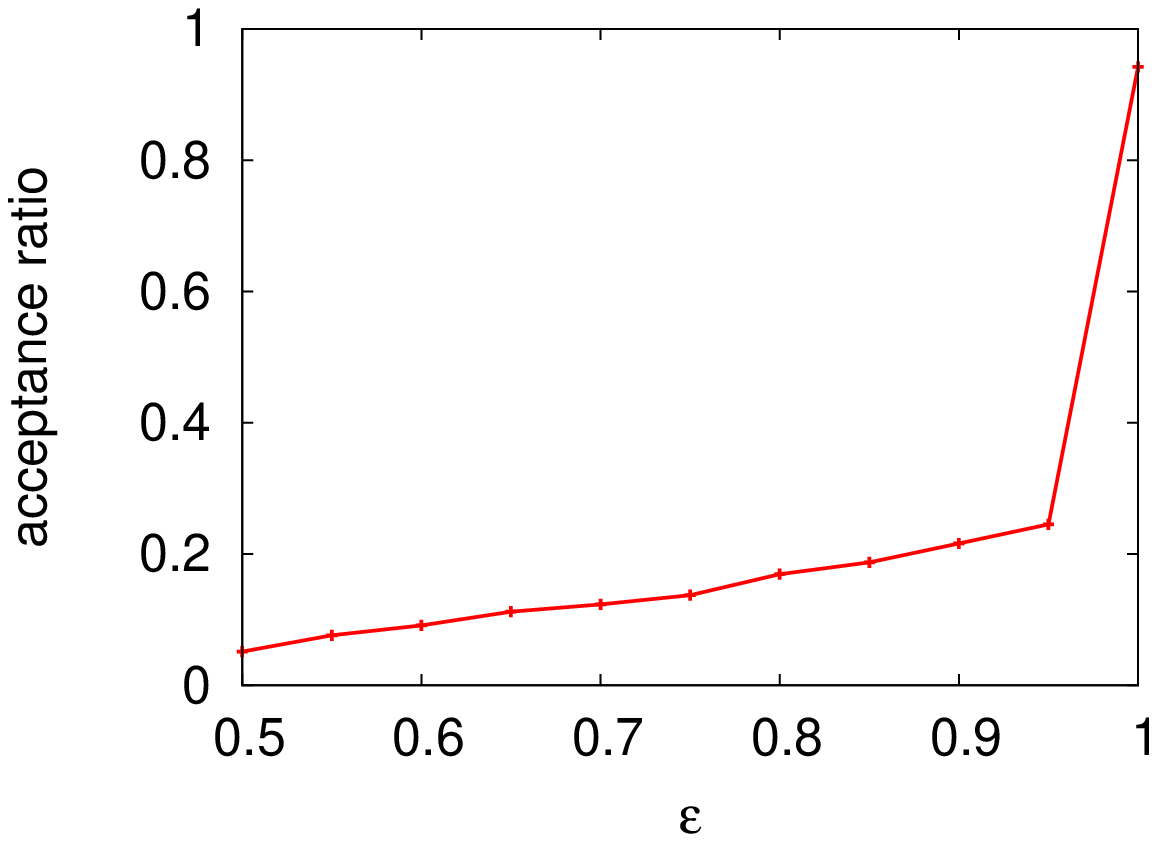} 
  \end{center}
\caption{Acceptance ratio (Eq.~\eqref{eq:Pacc_shoot}) as a function of the parameter $\epsilon$ in the Stoltz algorithm, Eq.~\eqref{eq:stoltz}, for a Markov chain of $1000$ trajectories, starting from the FCC basin at $T=0.12$, with trajectory lengths $\tau=500$ and a control parameter set to $\alpha=2000$.} 
  \label{fig:accept_stoltz}
\end{figure}

Values for the control parameters are chosen after observing the magnitude of the Lyapunov indicators $L(\mathbf{z})$ for few trajectories, and the difference between Lyapunov indicators $L(\mathbf{z})$ for current and trial trajectories in the shooting step (Eq.~\eqref{eq:Pacc_shoot}), in order to have an acceptance ratio for the shooting move not below $20\%$, see Fig.~\ref{fig:acceptance}. The choice of the trajectory length $\tau$ depends not on the Lyapunov bias, but on the necessity of having long enough trajectories to link the two funnels. $\tau$ needs also to be long enough to recover an appropriate statistics to compute reaction contants from correlation functions (see below). The use of the Stoltz algorithm (Eq.~\eqref{eq:stoltz}) in the shooting moves ensures that the energy distribution imposed by the preliminary MD is maintained along the simulation. The value for $\epsilon$ in Eq.~\eqref{eq:stoltz} is taken as $0.95$, so to have decorrelated sampled paths and ensure a sufficient acceptance ratio, see Fig.~\ref{fig:accept_stoltz}. 

We focus on the octahedral to icosahedral (FCC-ICO) transition at temperatures from $T=0.10$ to $T=0.15$: observing this passage using a direct MD or a standard TPS would require a considerable amount of CPU time (about $10^5$h, see Ref.~\cite{MP2007}) as the FCC configuration is at low temperatures the most stable one, so that the system rarely escapes from the FCC basin. In contrast, with our biased TPS technique we were able to observe at $T=0.12$ the first FCC-ICO reactive trajectories after about $300$ Markov chain steps.

To ensure that reactive paths start in the stable FCC state, we include in the path probability weight the constraining function (see Eq.~\eqref{costr_funct_1})
\begin{equation} \label{eq:phi_fcc}
\varphi^{FCC}(\mathbf{x}_{0})=\exp\left\{-\frac{\kappa}{2}\left(Q_{4}(\mathbf{x}_{0})-Q_{4}^{FCC}\right)^{2}\right\}
\end{equation}
assigned to the starting state $\mathbf{x}_0$ of the path, function of the bond order parameter $Q_{4}$ and centered on the value $Q_{4}^{FCC}=0.18$. We set $\kappa=500$, a sufficiently small stiffness that lets the trajectory starting point span the whole FCC basin. The function $\varphi^{FCC}$ keeps the beginning of the trajectories inside the FCC funnel, thus counterbalancing the effect of the local Lyapunov bias, that would pull trajectories on barriers.

\begin{figure}
  \begin{center}
\includegraphics{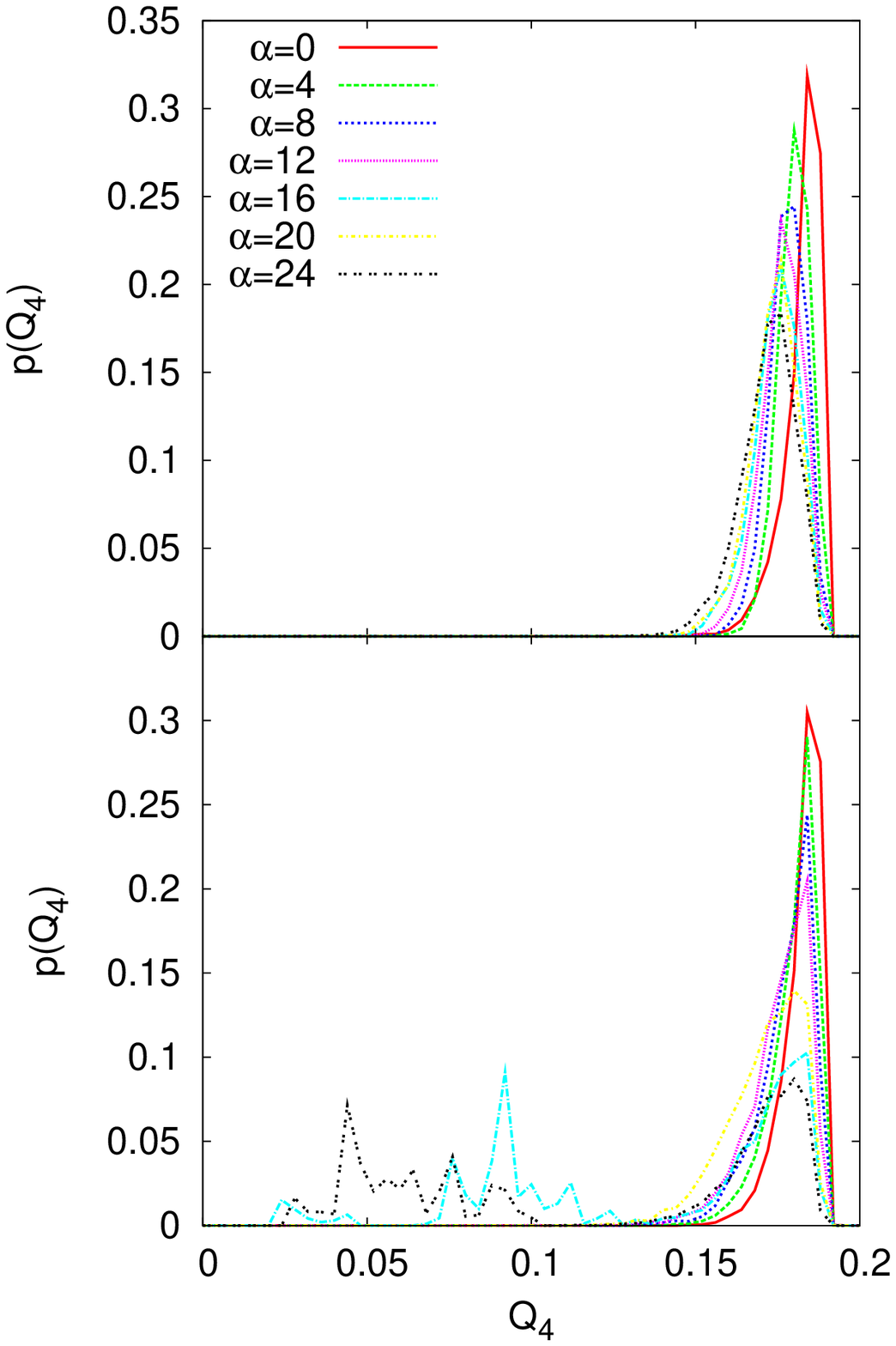}
  \end{center}
\caption{{\bf Top}: Histogram of the initial point position $\mathbf{x}_0$ for trajectories starting from the FCC basin at $T=0.12$ for different values of the control parameter $\alpha$, averaged on a Markov chain of 5000 steps. The restraining function of Eq.~\eqref{eq:phi_fcc} mantains the initial states of the trajectories in the FCC funnel for all $\alpha$ values. {\bf Bottom}: Same histogram, for the final position $\mathbf{x}_\tau$. Trajectories sampled with large $\alpha$ values escape the FCC funnel more often. Their final states are distributed over the whole FCC-ICO range.}
  \label{fig:histo_fc12}
\end{figure}

We present in Fig.~\ref{fig:histo_fc12} histograms for the first and the last point of the trajectories, for different values of the control parameter $\alpha$ at the FCC-ICO cohexistence temperature $T=0.12$. 
As $\alpha$ values increase, trajectories explore regions that are increasingly distant from the initial FCC basin, and some of them eventually cross the transition region and reach the ICO basin.  

Once reaction paths have been identified, the computation of the inter-funnel reaction constant by the correlation function via Eq.~\eqref{eq:c_kt} of Sec.~\ref{sec:Reaction-rate-constants-calculation} is possible if reactants and products basins are adjacent, i.e. if there is no intermediate state between them.~\cite{chandler1978statistical,dellago1999calculation} However, this hypothesis is not valid for the FCC-ICO transition: several results reported in the literature~\cite{DMW1999b,NCFD2000} show that reactive paths linking FCC and ICO states pass through many short-lived metastable basins, separated by barriers of different heights, not belonging to the two main funnels. These metastable states and transition regions have also been observed in a previous work using the transition current sampling method.~\cite{picciani2011simulating} Such a feature has been confirmed as well by an attentive analysis of our trajectories.

Among all the intermediate metastable states, we emphasize the presence of a basin related to a faulted FCC configuration, indicated with D in the following, having a bond order parameter value around  $Q_{4}=0.12$, already acknowledged in precedent studies.~\cite{AA2008,NCFD2000,picciani2011simulating} This basin has a rather important occupation probability if compared to other metastable states, and is visited by every reaction path linking FCC to ICO state. Moreover, this metastable state is indeed structurally related to the FCC basin, and the barrier separating the D structure from FCC is lower than the one separating the former from ICO state. As a result, several recrossing events of trajectories starting in FCC, visiting the D state and then going back to FCC, can be observed.

Hence, in order to correctly reconstruct the FCC to ICO transition paths, we have to take into account this intermediate metabasin. We therefore split  the FCC-ICO passage in two steps: the first part is given by the passage from the FCC basin to the D basin corresponding to $Q_{4}=0.12$. The second part is then given by trajectories starting from the D basin, and ending up in the ICO funnel.

To obtain this second part of FCC-ICO reactive paths, we constrain the first point of the trajectories to start in the D basin, using a constraining function given by an indicator on the bond-order parameter value $Q_{4}(\mathbf{x}_{0})$:
\begin{equation} \label{indic_q12}
h_{d}(Q_{4})=\begin{cases}
1 & 0.10\leq Q_{4}\leq0.13\\
0 & \rm{elsewhere}\end{cases}
\end{equation}

\begin{figure}
  \begin{center}
\includegraphics{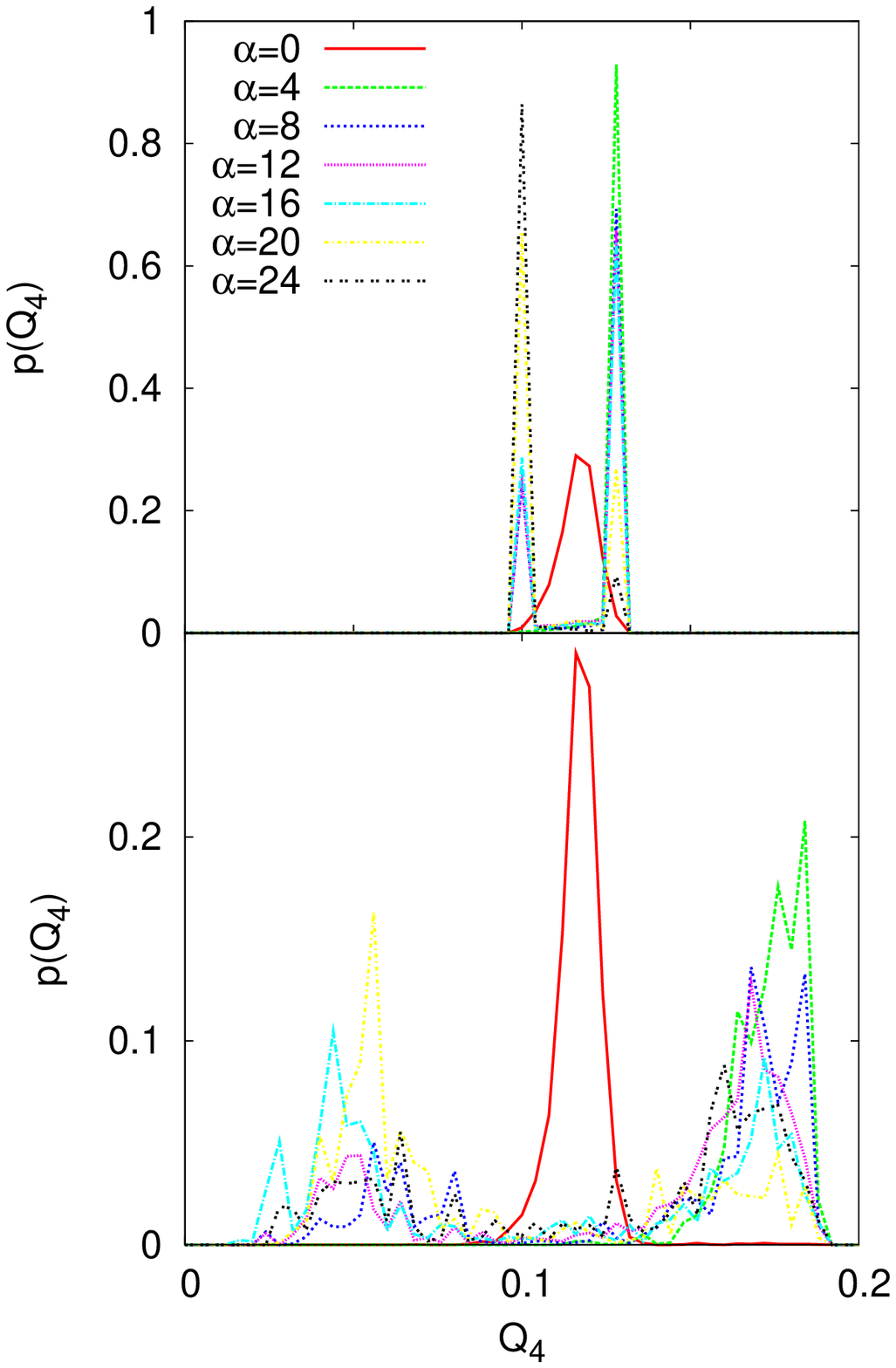}
  \end{center}
\caption{{\bf Top}: Histogram of the initial configuration $\mathbf{x}_0$ for trajectories starting from the D configuration metabasin located at $T=0.13$ for different values of the control parameter $\alpha$, averaged over a Markov chain of 5000 steps. {\bf Bottom}: Same histogram, for the final position $\mathbf{x}_\tau$. Trajectories end up in both the FCC or the ICO funnel.}
  \label{fig:histo_dc13}
\end{figure}

In Fig.~\ref{fig:histo_dc13} we present histograms for the distribution of the beginning and the end point of trajectories constrained with the indicator function of Eq.~\eqref{indic_q12}, at a temperature $T=0.13$ slightly above the solid-solid cohexistence. Paths sampled with the unbiased distribution $\alpha=0$ completely remain in the ``window'' given by $h_d(Q_{4}(\mathbf{x}_{0}))$. On the contrary, trajectories weighted with a Lyapunov bias tend to leave the metabasin: their starting points $\mathbf{x}_0$ tend to accumulate on the borders of the region defined by the indicator function in Eq.~\eqref{indic_q12}, while the end points $\mathbf{x}_\tau$ fall both in the FCC and the ICO funnels. 

In simulations performed at lower temperatures, this reconstruction of the second part of the FCC-ICO reactive path with trajectories starting from the D state is more difficult. Indeed, histograms for trajectories of the same length  at temperatures lower than the solid-solid transition $T=0.12$ show that an important fraction of the sampled trajectories
fall from the D state directly to the FCC basin, while a few trajectories end in the ICO state. This is attributed to the heights of the barriers separating the metastable D structure from either the stable ICO or stable FCC structures, the latter barrier being lower than the former one.

\subsubsection{FCC-ICO reaction constants}

The total reaction constants for the two-step FCC-ICO transition, assuming a steady occupation probability for the intermediate D state, is derived in Appendix~\ref{react_cost_interm} and reads 
\begin{equation} \label{eq:k_fi}
k_{F\rightarrow I}=\frac{k_{F\rightarrow d}k_{d\rightarrow I}}{(k_{d\rightarrow F}+k_{d\rightarrow I})}
\end{equation} where subscripts $F$, $d$ and $I$ refers to FCC, D and ICO states respectively.  The same steady state approximation is assumed for all intermediate metastable states in discrete path sampling studies.~\cite{W2003,W2002,TW2006}

Reaction rates $k_{F\rightarrow d}$ and $k_{d\rightarrow I}$ involve transitions between states separated by high free energy barriers,~\cite{picciani2011simulating} thus the hypothesis of time scale separation required by the reaction rate theory is still valid, and reaction constants can be computed using the method exposed in Sec.~\ref{sec:Reaction-rate-constants-calculation}. Reactive paths between FCC and D basins, and between D and ICO basins, are computed as reported above (Sec.~\ref{LJreactivepaths}). On the contrary, the D to FCC reaction rate $k_{d\rightarrow F}$ cannot be computed by LyTPS, because the requirement of a time scale separation is no longer valid, the barrier separating this two states being too low. It is therefore computed by direct MD simulation.

The reactivity $\mathcal A$  (Eq.~\eqref{eq:a_react}) for each trajectory is evaluated in simulations distinguishing the three basins FCC, ICO and D whose ranges of bond-order parameter $Q_4$ value, that is $0.13<Q_4<0.18$, $0<Q_4<0.04$ and $0.1<Q_4<0.13$, respectively. Data harvested during LyTPS runs are unbiased using MBAR.

\begin{figure}
  \begin{center}
\includegraphics[totalheight=9.5cm,width=20cm]{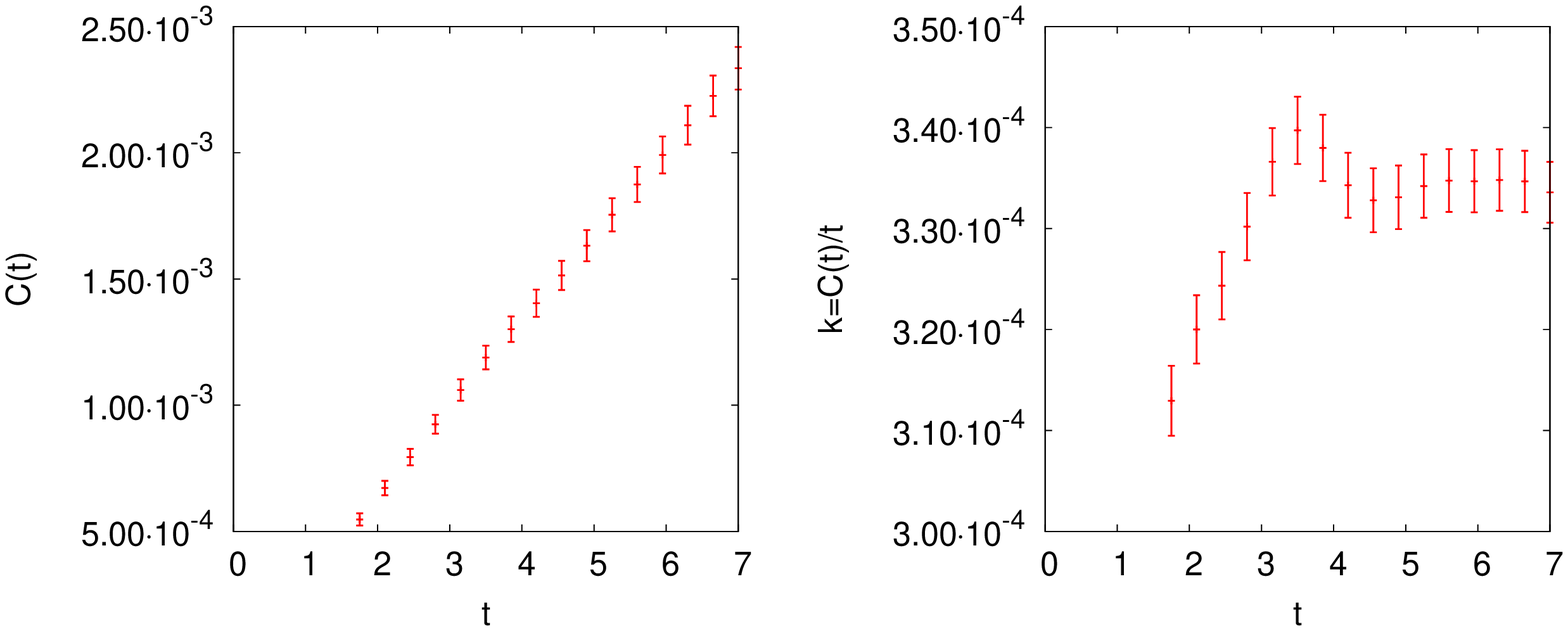}
  \end{center}
\caption{{\bf Left}: Correlation function for the transition from FCC to D basin, at $T=0.13$. Error bars are computed via Eq.~\eqref{estim} and directly given by Chodera's numerical recipe~\cite{chodera_site}. {\bf Right}: Reaction constant for this same passage, obtained at times shorter than the mean first passage time. The reactive flux $k(t)$ reaches a plateau value, corresponding to $k_{F\rightarrow d}$, as explained in Sec.~\ref{sec:Reaction-rate-constants-calculation}.}
  \label{fig:corr_fc13}
\end{figure}

\begin{figure}
  \begin{center}
 \includegraphics[totalheight=9.5cm,width=20cm]{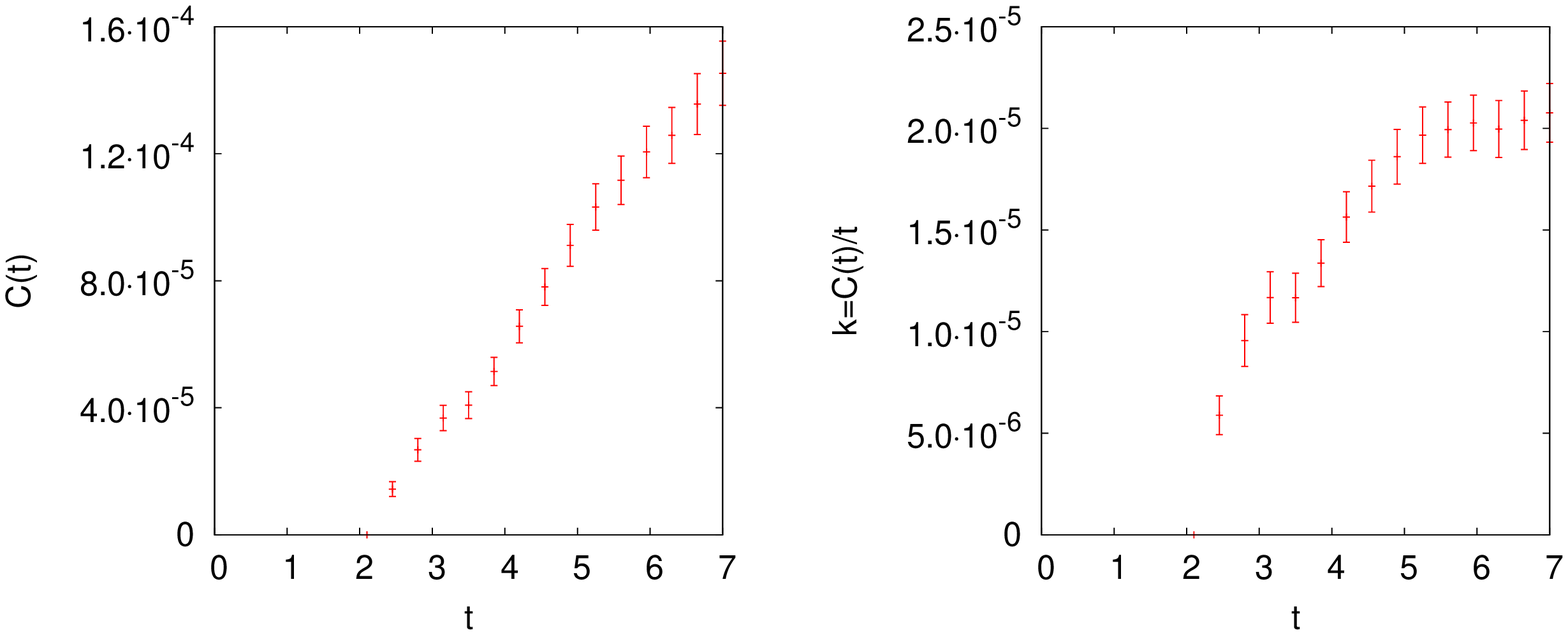}
  \end{center}
\caption{{\bf Top}: Correlation function for the D to ICO transition, at $T=0.13$ .  {\bf Bottom}: Reaction constant for this same passage.}
  \label{fig:corr_dc13}
\end{figure}

\begin{figure}[h!]
  \begin{center}
\includegraphics{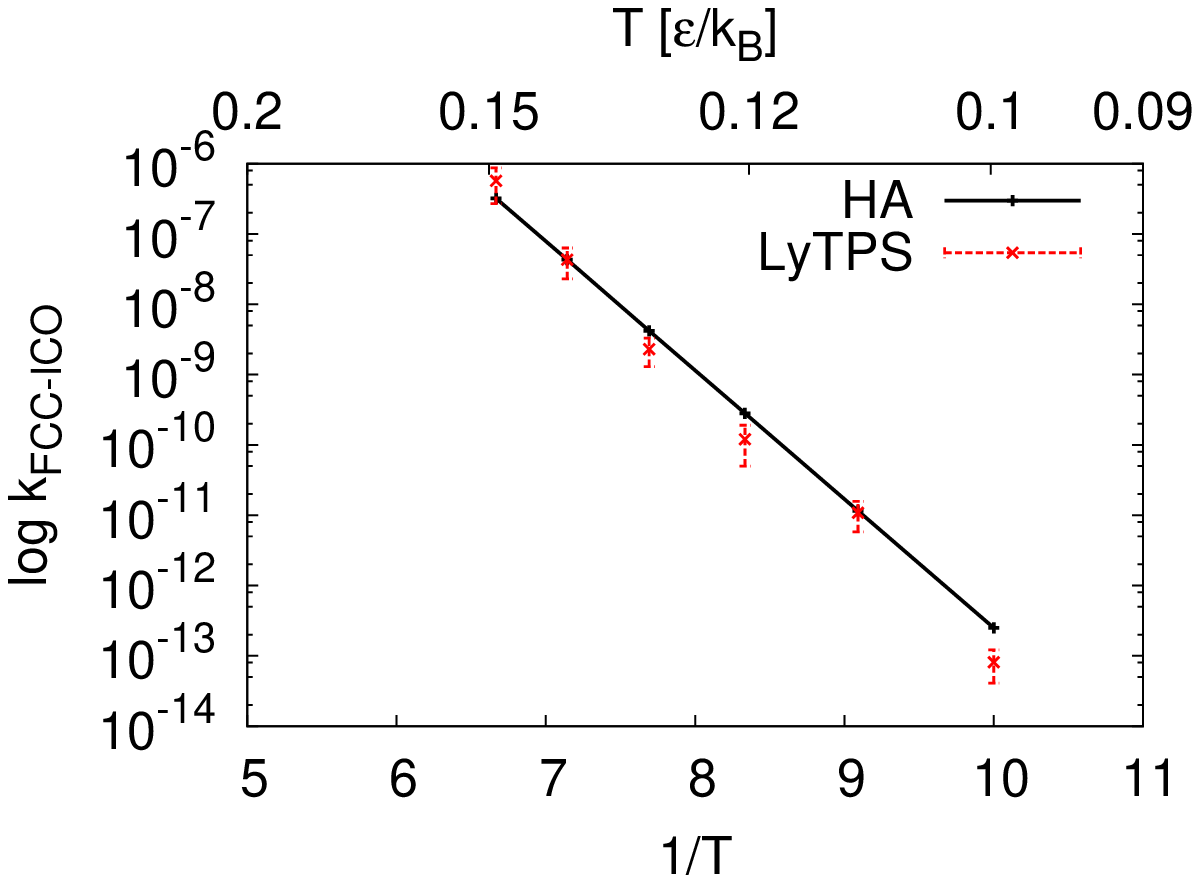} 
  \end{center}
\caption{Arrhenius plot for the FCC to ICO reaction rate from Table ~\ref{table} (LyTPS, red dots) compared with data obtained from DPS~\cite{W2002} using an harmonic approximation and reported in Ref.~\cite{W2003} (HA, black line). }
  \label{fig:arrhenius_lj38}
\end{figure}

In Fig.~\ref{fig:corr_fc13} and ~\ref{fig:corr_dc13}, two examples of population correlation functions for the computation of reaction rate constants, unbiased with MBAR, are reported. Note that reactivity values computed at short times are nearly zero, and do not contribute significatively to the correlation functions: in fact, these values are obtained from segments of trajectories too short to witness a complete transition between two states. In Table ~\ref{table}, we report reaction rate constants values for the FCC to D structure ($k_{F\rightarrow d}$) transition, and fro the D structure to ICO ($k_{d\rightarrow I}$), that give, through Eq.~\eqref{eq:k_fi}, a total FCC to ICO ($k_{F\rightarrow I}$) rate in good agreement with values given by DPS calculations~\cite{W2002,TW2006}. Finally, an Arrhenius plot comparing our results with the reaction constants proposed in Ref.~\cite{W2002,W2003} is presented in Fig.~\ref{fig:arrhenius_lj38}.

\begin{table}
\begin{tabular}{|c|c|c|c|c|}
\hline 
$T$ & $k_{F\rightarrow d}$ & $k_{d\rightarrow I}$ & $k_{F\rightarrow I}$ & $k_{F\rightarrow I}$ (Ref.~\cite{W2003})\tabularnewline
\hline
\hline 
0.10 & $1.2~10^{-7}$  & $1.4~10^{-7}$ & $8.1~10^{-14}$ & $2.5~10^{-13}$ \tabularnewline
\hline 
0.11 & $1.3~10^{-5}$  & $2.5~10^{-7}$ & $1.08~10^{-11}$ & $1.15~10^{-11}$ \tabularnewline
\hline 
0.12 & $8.1~10^{-5}$  & $4.0~10^{-7}$ & $1.2~10^{-10}$ & $2.82~10^{-10}$ \tabularnewline
\hline 
0.13 & $3.3~10^{-4}$  & $2.0~10^{-5}$ & $6.6~10^{-9}$ & $4.2~10^{-9}$ \tabularnewline
\hline 
0.14 & $9.3~10^{-4}$  & $4.5~10^{-5}$ & $4.3~10^{-8}$ & $4.3~10^{-8}$ \tabularnewline
\hline 
0.15 & $2.4~10^{-3}$ & $2.4~10^{-4}$ & $5.7~10^{-7}$ & $3.2~10^{-7}$ \tabularnewline
\hline
\end{tabular}
\caption{Table of reaction constants for the transitions from FCC to D structure, D to ICO, and the total FCC to ICO transition, indicated as $k_{F\rightarrow d}$, $k_{d\rightarrow I}$ and $k_{F\rightarrow I}$ respectively, at different temperatures. Values of $k_{F\rightarrow I}$ are obtained using Eq.~\eqref{eq:k_fi} and assuming the reaction constants $k_{d\rightarrow F}$ as $10^{-2}$, $3\cdot 10^{-2}$, $10^{-1}$ for $T=0.10$, $T=0.11$ and $T=0.12$ respectively (values obtained by Langevin MD), and unitary for $T\geq 0.12$. In the last column on the right, we report DPS data from Ref.~\cite{W2003}, computed in the harmonic approximation framework.}
\label{table}
\end{table}

\subsection{Vacancy migration in $\alpha$-Iron crystal}

The second example of a thermally activated process studied using LyTPS is the migration of a single vacancy in $\alpha$-Iron crystal. Atomic interactions of the model system are described by an embedded atom model potential.~\cite{ackland2004} The crystal structure is body-centered cubic, and the initial unrelaxed cell contains 1023 atoms displayed on 1024 lattice sites, the vacant site corresponding to the vacancy. The reaction coordinate used to represent the motion of the vacancy is the distance crossed by the moving atom that replaces the vacancy.

The free energy landscape for this system has been reconstructed in Ref.~\cite{athenes2010free}. It presents two stable states, the first corresponding to the initial configuration, and the second one to the same configuration obtained by translating a neighboring atom into the vacancy site. The first neighbor distance is $a=2.47 \textrm{\AA}$,  switching its initial position with the vacancy site. There is a single free energy barrier separating these two states for temperatures above $T=450 K$, while for lower temperatures an intermediate metastable state appears, corresponding to an intra-site position for the moving atom.~\cite{malerba2010comparison}

We performed LyTPS simulations with trajectories of different lengths (see below), with time step $\delta t=4\cdot10^{-15} s$. A preliminary MD simulation is done to equilibrate the system to the required temperature, with a friction parameter $\gamma =2.5\cdot 10^{12} s^{-1}$. We explored temperatures ranging from $300 K$ to $850 K$. The LyTPS shooting and shifting moves are iterated to cosntruct Markov chains consisting of $1500$ trajectories.

As for LJ$_{38}$, the trajectory length and the values for the control parameter $\alpha$ have been chosen in order to ensure an acceptance ratio of $25\%$ and an adequate ergodic sampling of the phase space. For temperatures above $450 K$, the presence of a single "smooth" barrier separating the two metastable states makes this application simple enough: sampling of reactive trajectories is achieved using $15$ $\alpha$ values from the unbiased simulation at $\alpha=0$ up to $\alpha=150\cdot 10^{12}$, with trajectories of $300$ steps. 
For temperatures below $450K$, an ergodic sampling of trajectory space appears more difficult. We therefore employed longer trajectories of $500$ time steps, as well as larger values of the control parameter, up to $\alpha=500\cdot 10^{12}$ to allow the system to escape the initial basin.

Reaction constants for the passage between the two stable states above $450K$ are estimated from correlation functions unbiased with the MBAR algorithm, via Eq.~\eqref{eq:c_kt}.

For $T<450K$, the presence of an intermediate metastable basin has to be taken into account in the evaluation of reaction constants. As recalled in Sec.~\ref{lj38}, the reaction rate expression obtained from Eq.~\eqref{eq:c_kt} holds only for adjacent reactant and product basins. At low temperatures, it is therefore more appropriate to use our algorithm to evaluate the reaction constant for the passage from the initial state to the intermediate basin. From Eq.~\eqref{eq:k_fi}, reaction rate for the passage from one stable configuration to the other is simply half the reaction constant from one stable configuration to the intermediate one. Indeed, reaction constants for transitions from the intermediate metastable state to either of the two stable states are equal, because of the symmetric shape of the potential surface.~\cite{malerba2010comparison} 

\begin{figure}
  \begin{center}
\scalebox{0.8}{\includegraphics{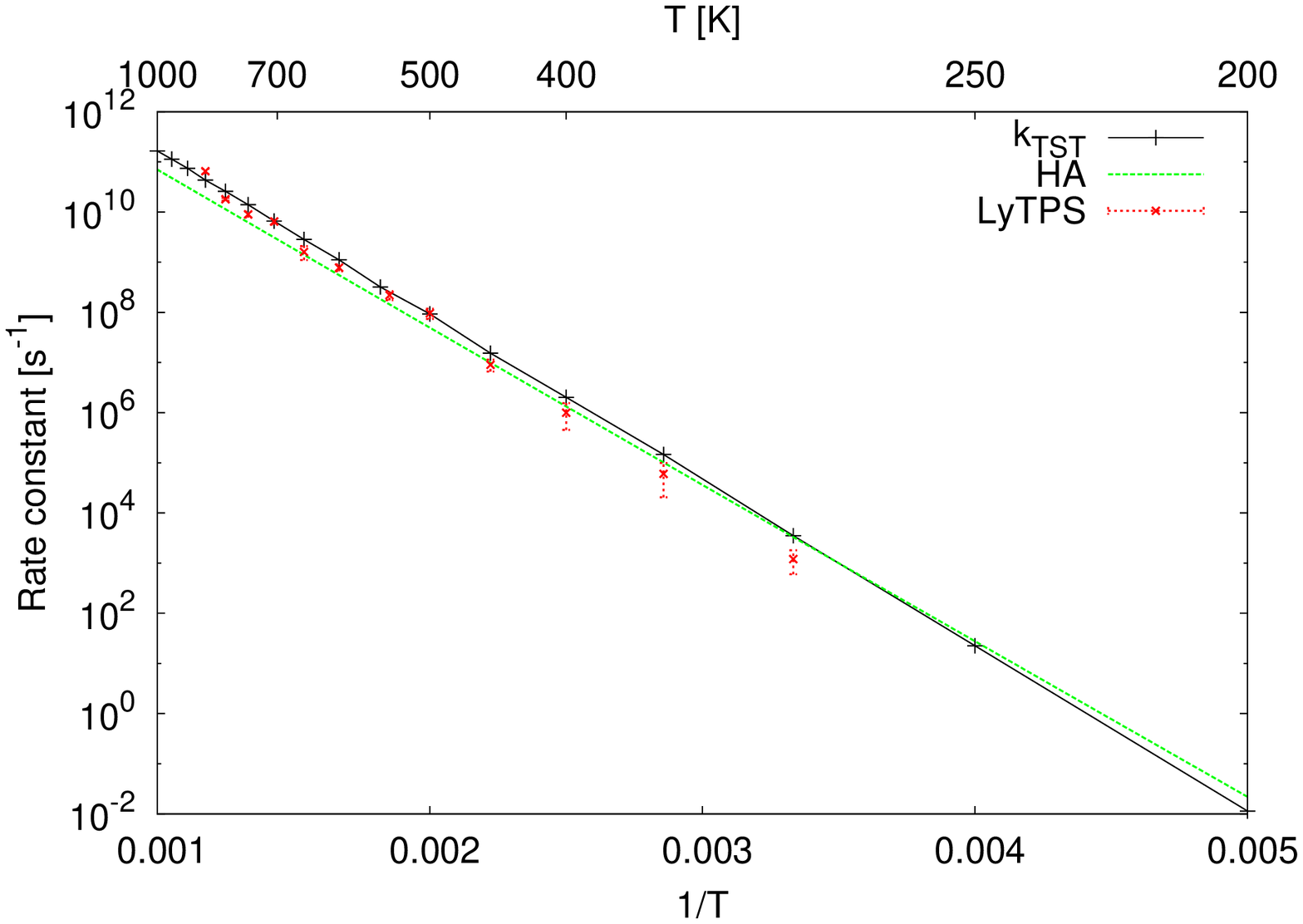}}
  \end{center}
\caption{Arrhenius plot of reaction constants for migration of monovacancy in $\alpha$-Iron obtained with Lyapunov biased TPS (LyTPS, red points), compared with rates obtained using in Eq.~\eqref{kappa_tst} the free energy barriers proposed in Ref.~\cite{athenes2010free} ($k_{TST}$, black line) and using an harmonic approximation (HA, green line).}
  \label{fig:arrhenius_vac}
\end{figure}

In Fig.~\ref{fig:arrhenius_vac} we compare the reaction rates obtained with LyTPS, those computed inserting in the transition state theory (TST) expression (Eq.~\eqref{kappa_tst}) the free energy barriers reported in Ref.~\cite{athenes2010free}, and reaction constants obtained with a classical harmonic approximation (HA).
Above the Debye temperature ($470K$), rates obtained with LyTPS fall between TST and harmonic approximation values. To explain this point, we recall that reaction rates we estimate with the method exposed in Sec.~\ref{sec:Reaction-rate-constants-calculation} are derived from Eq.~\eqref{eq:c_kt}, hence correspond to the \textit{phenomenological} rate constants. These values are therefore bounded from above by TST values, that overestimate reaction rates~\cite{chandler1978statistical}, as can be seen from Eq.~\eqref{eq:react_flux}. Conversely, values obtained with the harmonic approximation neglect anharmonicity effects on the activation barrier, thus giving reaction rates that are lower then the phenomenological rate constants we compute. Our results are then in agreement with the reaction rate theory recalled in Sec.~\ref{sec:Reaction-rate-constants-calculation}.

\section{Conclusion}

The method presented in this paper allows to compute reaction rate constants for inter funnel transitions in many-body systems. The reaction rate values are evaluated using a path sampling algorithm biased with local Lyapunov numbers. This bias aims at enhancing an accelerate sampling of reactive paths, so as to reduce the lenght of Markov chains and the amount of CPU time to observe activated processes. We assess the performace of the method by observing reaction paths and evaluating equilibrium reaction rates for structural transitions in the LJ$_{38}$ system and for vacancy migration in an $\alpha$-Iron crystal. For both systems, we were able to predict phenomenological rate constants, in very good agreement with data already given in the literature in the case of LJ$_{38}$.

The Lyapunov biased TPS method presents several advantages, and incorporates features of different rare events simulation methods.

Firstly, with respect to other importance sampling methods based on Lyapunov weighted sampling~\cite{TK2006,geiger2010identifying}, Lyapunov biased TPS has the main advantage of a simpler implementation. This is due to the Lyapunov indicator $L(\mathbf z)$ we propose in Eq.~\eqref{eq:15}, that allows to quantify the divergence of hamiltonian trajectories by resorting to local Lyapunov numbers. These quantities can be easily calculated with the Lanczos algorithm, that enables one to compute the largest eigenvalues of the Jacobian matrix of the hamiltonian mapping with a limited computational cost. As recalled in Sec.~\eqref{rliecc}, resorting to local Lyapunov numbers to evaluate chaoticity of phase space trajectories doesn't suffer from the computational drawbacks of other algorithms aimed at the same purpose, as RLI or the tangent space method.~\cite{geiger2010identifying} The implementation of shooting and shifting Monte Carlo moves in a Lyapunov biased TPS is less complicated, and computationally less expensive, than the algorithm proposed in Ref.~\cite{geiger2010identifying} with the use of RLI, because we do not need to compute four trajectories to evaluate the divergence of a single path~\cite{sandor2004relative,geiger2010identifying}.

Secondly, this formulation for the Lyapunov indicator is such that the bias applied to each path in order to enhance the fraction of reactive trajectories is clearly identified, contrarily to what is done in rather complex cloning algorithms like the one proposed in Lyapunov weighted dynamics~\cite{TK2006} and transition current sampling~\cite{picciani2011simulating}. Hence, the use of standard unbiasing statistical tools to recover unbiased observables is possible and requires a small theoretical and computational effort. 

Furthermore, we consider the access to the evaluation of equilibrium transition rates as the most important aspect of Lyapunov biased TPS. On the computational point of view, the direct access to reaction rates without resorting to a distinct evaluation of the reaction barriers and the transmission factor, as usually done in standard TPS technique~\cite{DBCD1998}, is a very advantageous feature. To unbias reaction constants computed in Lyapunov biased ensembles we chose among other unbiasing algorithm, like WHAM~\cite{wham} or Extended Bridge Sampling~\cite{ebs} techniques, the MBAR method~\cite{shirts2008statistically}. MBAR has proven to be computationally efficient and to give an adequate numerical precision in estimating reaction constants. Moreover, this work is the first in which MBAR is implemented exploiting the marginal probability derived from a waste recycling method. This allows to include informations encoded in rejected proposal trajectories.

Finally, LyTPS is implemented at a finite temperature, imposed to trajectories by the canonical distribution from which the path starting point is selected and maintained along the path thanks to the Stoltz proposal for the shooting algorithm, see Sec.~\ref{sec:Transition-path-sampling}. In parallel, the Lyapunov indicator used as a bias to select reactive paths directly links the path sampling to the local conformation of the potential energy surface via the hessian matrix, thus giving to our method an intrinsic dependence on the potential energy landscape. The coupling between a finite-temperature sampling and potential energy surface conformation is a noticeable improvement if compared to eigenvector-following methods, that are based on the shape of the potential energy surface, but usually operate at zero temperature. LyTPS can be acknowledged as a finite, nonzero temperature version of the well-known eigenvector-following techniques, such as Dimer, Optim or ART.~\cite{PhysRevLett.77.4358,PhysRevE.57.2419}

The advantages related to a finite temperature technique do not concern only the exploration of the energy landscape, but also the fact that the evaluation of physical observables like reaction rates takes into account temperature and anharmonicity effects. Indeed, the phenomenological reaction rate we computed (see Sec.~\ref{sec:Reaction-rate-constants-calculation}) can be compared with experimental measures: LyTPS turns out to be a powerful tool to study problems like vacancy migration in Iron. In this context, reaction rates are usually estimated using only the \textit{potential energy} barriers at 0 K and harmonic approximations: this give approximate results with respect to experimental data obtained at nonzero temperatures. LyTPS reaction rates find an important application as input parameters for object Monte Carlo codes aimed at numerically reproducing resistivity recovery experiments. 

We conclude observing that this method can be furhter developed using a parallel tempering technique, or implemented for the computation of reaction rates in more complex condensed matter systems, and can find interesting applications in a wide class of research fields, spanning from molecular biophysics to  physical metallurgy, where the numerical determination of reaction rates has important consequences for experimental applications.

\section{Acknowledgements}

I warmly thank Manuel Athenes and Mihai-Cosmin Marinica for help and fruitful discussions.

\appendix

\section{Reaction constants for processes with intermediate states} \label{react_cost_interm}
The time evolution of the occupation probabilities ${p}_{F}(t)$, ${p}_{I}(t)$ and ${p}_{d}(t)$ of the FCC, ICO and faulted states  respectively, read 
\begin{equation}
 \begin{cases}
\dot{p}_{F}(t)= & -k_{F\rightarrow d}p_{F}(t)+k_{d\rightarrow F}p_{d}(t)\\
\dot{p}_{d}(t)= & k_{F\rightarrow d}p_{F}(t)-(k_{d\rightarrow F}+k_{d\rightarrow I})p_{d}(t)+k_{I\rightarrow d}p_{I}(t)\\
\dot{p}_{I}(t)= & k_{d\rightarrow I}p_{d}(t)-k_{I\rightarrow d}p_{I}(t)\end{cases}
\end{equation}
where subscripts $F$, $d$ and $I$ refers to FCC, faulted and ICO states respectively, and $k_{A\rightarrow B}$ indicates the generic transition rate from state A to state B. Note that the possibility of a direct transition from FCC to ICO states without passing by the intermediate basin has been neglected. The stationary approximation for the metastable state imposes $\dot{p}_{d}(t)=0$, thus the above system can be recasted in the form  \[
\begin{cases}
\dot{p}_{F}(t)= & -\frac{k_{F\rightarrow d}k_{d\rightarrow I}}{(k_{d\rightarrow F}+k_{d\rightarrow I})}p_{F}(t)+\frac{k_{I\rightarrow d}k_{d\rightarrow F}}{(k_{d\rightarrow F}+k_{d\rightarrow I})}p_{I}(t)\\
\dot{p}_{I}(t)= & \frac{k_{F\rightarrow d}k_{d\rightarrow I}}{(k_{d\rightarrow F}+k_{d\rightarrow I})}p_{F}(t)-\frac{k_{I\rightarrow d}k_{d\rightarrow F}}{(k_{d\rightarrow F}+k_{d\rightarrow I})}p_{I}(t)\end{cases}\]
and the reaction constant between FCC and ICO states can be calculated as 
\begin{equation} \label{eq:k_fi}
k_{F\rightarrow I}=\frac{k_{F\rightarrow d}k_{d\rightarrow I}}{(k_{d\rightarrow F}+k_{d\rightarrow I})}.
\end{equation}

\bibliographystyle{elsarticle-num}

\begin{thebibliography}{99}

\bibitem{W2003}
D.~Wales, {Energy landscapes}, Springer, 2003.

\bibitem{DBCD1998} C.~Dellago, P.~Bolhuis, F.~Csajka, D.~Chandler, J. Chem. Phys. \textbf{108} (1998) 1964.

\bibitem{BCDG2002} P.~Bolhuis, D.~Chandler, C.~Dellago, P.~Geissler, Annu. Rev. Phys. Chem. \textbf{53} (2002) 291. 







\bibitem{MB1998}
N.~Mousseau, G.~Barkema, Phys. Rev. E \textbf{57} (1998) 2419.


\bibitem{dellagohooverposch} C.~Dellago, W.G. Hoover, H.A. Phys. Rev. E \textbf{53} (1996) 1485


\bibitem{TTK2006}
J.~Tailleur, S.~T{\u{a}}nase-Nicola, J.~Kurchan, J. Stat. Phys. \textbf{122} (2006) 557.

\bibitem{TK2004}
S.~T{\u{a}}nase-Nicola, J.~Kurchan, J. Stat. Phys. \textbf{116} (2004) 1201.

\bibitem{TK2006}
J.~Tailleur, J.~Kurchan, Nature Phys.~\textbf{3} (2007) 203. 

\bibitem{TK2003b}
S.~T{\u{a}}nase-Nicola, J.~Kurchan, Phys. Rev. Lett. \textbf{91} (2003) 188302.

\bibitem{HTB1990} P.~H{\"a}nggi, P.~Talkner, M.~Borkovec, Rev. Mod. Phys. \textbf{62} (1990) 251. 


\bibitem{AAC2006}
G.~Adjanor, M.~Ath{\`e}nes, F.~Calvo, Eur. Phys. J. B \textbf{53} (2006) 47.

\bibitem{AA2008}
M.~Ath{\`e}nes, G.~Adjanor, J. Chem. Phys. \textbf{129} (2008) 024116.

\bibitem{C2007}
W.~Cochran, {Sampling techniques}, Wiley India Pvt. Ltd., 2007. 

\bibitem{DMW1999b}
J.~Doye, M.~Miller, D.~Wales, J. Chem. Phys. \textbf{111} (1999) 8417. 

\bibitem{DMW1999}
J.~Doye, M.~Miller, D.~Wales, J. Chem. Phys. \textbf{110} (1999) 6896.


\bibitem{shimada}
I.~Shimada, T.~Nagashima, Prog. Theor. Phys. \textbf{61} (1979) 1605.

\bibitem{NCFD2000}
J.~Neirotti, F.~Calvo, D.~Freeman, J.~Doll, J. Chem. Phys. \textbf{112} (2000) 10340.

\bibitem{CNFD2000}
F.~Calvo, J.~Neirotti, D.~Freeman, J.~Doll, J. Chem. Phys. \textbf{112} (2000) 10350.

\bibitem{MF2006}
V.~Mandelshtam, P.~Frantsuzov, J. Chem. Phys. \textbf{124} (2006) 204511.

\bibitem{BWC2006}
T.~Bogdan, D.~Wales, F.~Calvo, J. Chem. Phys. \textbf{124} (2006) 044102.

\bibitem{PCABD2006} P.~Poulain, F.~Calvo, R.~Antoine, M.~Broyer, P.~Dugourd, Phys. Rev. E \textbf{73} (2006) 056704.

\bibitem{MP2007} T.~Miller, C.~Predescu,  J. of Chem. Phys. \textbf{126} (2007) 144102.

\bibitem{AM2010} M.~Ath\`{e}nes, M.-C.. Marinica, J. Comput. Phys. \textbf{229} (2010) 7129.

\bibitem{W2002} D.~Wales, Mol. Phys. \textbf{100} (2002) 3285.

\bibitem{W2004} D.~Wales, Mol. Phys. \textbf{102} (2004) 891. 

\bibitem{BW2006} D.~J. Wales, J. Chem. Phys. \textbf{124} (2006) 234110. 

\bibitem{meyer} H.~D. Meyer, J. Chem. Phys. \textbf{84} (1986) 3147. 

\bibitem{TW2006} S.~Trygubenko, D. Wales, J. Chem. Phys. \textbf{124} (2006) 234110. 

\bibitem{DBC1998} C.~Dellago, P. Bolhuis, D.~Chandler, J. Chem. Phys.~\textbf{108} (1998) 9236.

\bibitem{SNR1981} P.~Steinhardt, D.~Nelson, M.~Ronchetti, Phys. Rev. Lett. \textbf{47} (1981) 1297.

\bibitem{SNR1983} P.~Steinhardt, D.~Nelson, M.~Ronchetti, Phys. Rev.B \textbf{28} (1983) 784. 

\bibitem{ZCL2007} L.~Zhan, J.~Z.~Y. Chen, W.~Liu, J. Chem. Phys. \textbf{127} (2007) 141101. 



\bibitem{geiger2010identifying} P. Geiger, C. Dellago, Chem. Phys.~\textbf{375} (2010) 309. 



\bibitem{ott2002chaos} Ott, E.,Cambridge Univ. Press (2002)

\bibitem{marinica2011energy} M.C. Marinica,  F. Willaime, N. Mousseau, N., Phys. Rev. B vol.9, \textbf{83} (2011) 094119

\bibitem{dellago2002transition}  C. Dellago, P.G. Bolhuis, P.L. Geissler, Wiley Online Library (2002)

\bibitem{PhysRevLett.77.4358} G. T. Barkema, N. Mousseau, Phys. Rev. Lett. \textbf{77} 4358 (1996)

\bibitem{PhysRevE.57.2419} G. T. Barkema, N. Mousseau, Phys. Rev. E \textbf{57} 2419 (1998)

\bibitem{calvo1998chaos} F. Calvo, J. Chem. Phys. \textbf{108} 6861 (1998)

\bibitem{hinde1992chaos} R.J. Hinde, R.S. Berry, D.J. Wales, J. Chem. Phys. \textbf{96} 1376 (1992) 

\bibitem{hinde1993chaotic} R.J. Hinde, R.S. Berry, J. Chem. Phys. \textbf{99} 2942 (1993) 

\bibitem{ruelle1978thermodynamic} D. Ruelle, G. Gallavotti, {Thermodynamic formalism} Addison-Wesley Reading (1978)

\bibitem{sandor2004relative} Z. S{\'a}ndor, B. {\'E}rdi, A. Sz{\'e}ll, B. Funk, Celestial Mechanics and Dynamical Astronomy \textbf{90} 127 (2004)

\bibitem{lanczos1961applied} C. Lanczos {Applied analysis} Prentice Hall (1961)

\bibitem{dellago1999calculation} C. Dellago,P.G. Bolhuis, D. Chandler, J. Chem. Phys. \textbf{110} 6617 (1999) 
 
\bibitem{pratt1986statistical} L.R. Pratt, J. Chem. Phys. \textbf{85} 5045 (1986) 

\bibitem{benettin1980lyapunov} G. Benettin, L. Galgani, A. Giorgilli,  J.M. Strelcyn, Meccanica \textbf{15} 9 (1980)

\bibitem{shirts2008statistically} M.R. Shirts, J.D. Chodera, J. Chem. Phys. \textbf{129} 124105 (2008) 

\bibitem{oseledec1968multiplicative} V.I. Oseledec, Trans. Moscow Math. Soc \textbf{19} 197 (1968)
 
\bibitem{lichtenberg1992regular} A.J. Lichtenberg, M.A. Lieberman, {Regular and chaotic dynamics} Springer Verlag (1992)

\bibitem {dellagoposch} C. Dellago, W.G. Hoover, H.A.Posch, Phys. Rev. E \textbf{65} 056216 (2002)

\bibitem{chandler1987introduction} D. Chandler, {Introduction to modern statistical mechanics} Oxford Univ. Press (1987)
 
\bibitem{frenkel2002understanding} D. Frenkel, B. Smit, {Understanding molecular simulation: from algorithms to applications} Academic Press (2002)

\bibitem{damask1963point} A.C.Damask, G.J. Dienes, {Point defect in metals} Gordon and Breach (1963)

\bibitem{stoltz2007path} G.Stoltz, J. Comp. Phys. (2007)

\bibitem{cosminart} N. Mousseau, L. Karim Bland, P. Brommer, J.F. Joly, F. El-Mellouhi, E. Machado-Charry, M.C. Marinica, P. Pochet, preprint


\bibitem{chandler1978statistical} D. Chandler, J. Chem. Phys. \textbf{68} 2959 (1978)

\bibitem{minh2009optimal} D.D.L.Minh,J.D. Chodera, J. Chem. Phys. \textbf{131} 134110 (2009)
 
\bibitem{bennett1976efficient} C.H. Bennett, J. Comp. Phys. \textbf{22} 245 (1976)
  
\bibitem{amitrano1992probability} C. Amitrano, R.S. Berry, Phys. Rev. Lett. \textbf{68} 729 (1992)

\bibitem{wales2009calculating} D.J. Wales, J. Chem. Phys. \textbf{130} 204111 (2009)
  
\bibitem{evans1990viscosity} D.J. Evans, EGD Cohen, G.P. Morriss, Phys. Rev. A \textbf{42} 5990 (1990)

\bibitem{butera1987phase} P. Butera, G. Caravati, Phys. Rev. A \textbf{36} 962 (1987)
 
\bibitem{paladin1986scaling} G. Paladin, A. Vulpiani, J. Phys. A \textbf{19} 1881 (1986)
 
\bibitem{benettin1976kolmogorov} G. Benettin, L. Galgani, J.M. Strelcyn, Phys. Rev. A \textbf{14} 2338 (1976)

\bibitem{kolmogorov1959entropy} A.N. Kolmogorov, Dokl. Acad. Nauk. \textbf{124} 754 (1959)

\bibitem{pesin1977characteristic} Y.B. Pesin, Russian Mathematical Surveys \textbf{32} 55 (1977)

\bibitem{abarbanel1992} H. D. I. Abarbanel, R. Brown, M.B. Kennel, J. Nonlinear Sci. \textbf{2} 343 (1992)

\bibitem{adjanor2011waste} G. Adjanor,M. Ath{\`e}nes, J.M. Rodgers, J. Chem. Phys. \textbf{135} 044127 (2011)

\bibitem{wr1} J.F. Delmas, B. Jourdain, Arxiv preprint math/0611949 (2006)

\bibitem{wr2} J. Kim, J. Rodgers, M. Athenes, B. smit, J. Chem. Theory Comput. (2011) 

\bibitem{picciani2011simulating} M. Picciani, M. Ath{\`e}nes, J. Kurchan, J. Tailleur, J. Chem. Phys. \textbf{135} 034108 (2011)

\bibitem{athenes2010free} Ath{\`e}nes, M. and Marinica, M.C., J. Comp. Phys. \textbf{229} 7129 (2010)

\bibitem{ackland2004} G. Ackland, M. Medelev, D. Srolovitz, S. Han, A. Barashev, J. Phys.: Condens. Matter \textbf{16} 2629 (2004)

\bibitem{wham} S. Kumar, D. Bouzida, R. H. Swendsen, P. A. Kollman, J. M. Rosenberg, J. Comp. Chem. \textbf{13} 1011 (1992)

\bibitem{ebs} Z. Tan, J. Am. Stat. Assoc. \textbf{99} 1027 (2004)

\bibitem{chodera_site} https://simtk.org/home/pymbar

\bibitem{malerba2010comparison} L. Malerba, M.C. Marinica, N. Anento, C. Bjorkas, H. Nguyen, C. Domain, F. Djurabekova, P. Olsson, K. Nordlund, A. Serra, D. Terentyev, F. Willaime, C.S. Becquart, J. Nucl. Mat. \textbf{406} 19 (2010)

\end{thebibliography}

\end{document}